\documentclass[aps,prx,twocolumn,superscriptaddress]{revtex4-2}

\usepackage{graphicx}

\newcommand{\upperRomannumeral}[1]{\uppercase\expandafter{\romannumeral#1}}

\usepackage{hyperref}
\pdfoutput=1
\usepackage{comment}
\usepackage{multirow}
\usepackage{array}
\usepackage{amsmath}
\usepackage{bm}
\usepackage{amssymb}
\usepackage{textcomp}
\usepackage{stmaryrd} 
\usepackage{ulem}
\usepackage{gensymb}

\usepackage{import}
\usepackage{color}
\usepackage{verbatim}

\usepackage{float}

\usepackage{lipsum}
\usepackage{sistyle} 

\newcommand{\wDD}{\omega_{\mathrm{2D}}}
\newcommand{\wG}{\omega_{\mathrm{G}}}
\newcommand{\GammaG}{\Gamma_{\mathrm{G}}}
\newcommand{\GammaDD}{\Gamma_{\mathrm{2D}}}

\renewcommand{\a}{\alpha}
\renewcommand{\b}{\beta}
\renewcommand{\k}{\mathbf{k}}
\newcommand{\q}{\mathbf{q}}
\renewcommand{\r}{\mathbf{r}}
\renewcommand{\l}{\lambda}
\newcommand{\pd}{\partial}
\newcommand{\ppsi}{\hat{\psi}}
\newcommand{\ppsidag}{\hat{\psi}^{\dagger}}
\newcommand{\intr}{\int \mathrm{d}^{3} r}
\newcommand{\intt}{\int \mathrm{d} t}

\def\XXint#1#2#3{{\setbox0=\hbox{$#1{#2#3}{\int}$}
     \vcenter{\hbox{$#2#3$}}\kern-.5\wd0}}

\newcommand{\wn}{\mathrm{cm}^{-1}}

\begin{document}

\title{Symmetry-dependent dielectric screening of optical phonons in monolayer graphene}

\author{Loïc Moczko}
\affiliation{Universit\'e de Strasbourg, CNRS, Institut de Physique et Chimie des Mat\'eriaux de Strasbourg, UMR 7504, F-67000 Strasbourg, France}

\author{Sven Reichardt}
\affiliation{Department of Physics and Materials Science, University of Luxembourg, 162a avenue de la Fa\"iencerie, L-1511 Luxembourg, Luxembourg }

\author{Aditya Singh}
\affiliation{Universit\'e de Strasbourg, CNRS, Institut de Physique et Chimie des Mat\'eriaux de Strasbourg, UMR 7504, F-67000 Strasbourg, France}
\affiliation{Department of Physics, Indian Institute of Technology Delhi, Hauz Khas, New Delhi 110016, India}

\author{Xin Zhang}
\altaffiliation{Present address: State Key Laboratory of Superlattices and Microstructures, Institute of Semiconductors, Chinese Academy of Sciences, Beijing 100083, China}
\affiliation{Universit\'e de Strasbourg, CNRS, Institut de Physique et Chimie des Mat\'eriaux de Strasbourg, UMR 7504, F-67000 Strasbourg, France}

\author{Luis E. Parra L\'opez}
\affiliation{Universit\'e de Strasbourg, CNRS, Institut de Physique et Chimie des Mat\'eriaux de Strasbourg, UMR 7504, F-67000 Strasbourg, France}

\author{Joanna L. P. Wolff}
\affiliation{Universit\'e de Strasbourg, CNRS, Institut de Physique et Chimie des Mat\'eriaux de Strasbourg, UMR 7504, F-67000 Strasbourg, France}

\author{Aditi Raman Moghe}
\affiliation{Universit\'e de Strasbourg, CNRS, Institut de Physique et Chimie des Mat\'eriaux de Strasbourg, UMR 7504, F-67000 Strasbourg, France}

\author{Etienne Lorchat}
\affiliation{Universit\'e de Strasbourg, CNRS, Institut de Physique et Chimie des Mat\'eriaux de Strasbourg, UMR 7504, F-67000 Strasbourg, France}

\author{Rajendra Singh}
\affiliation{Department of Physics, Indian Institute of Technology Delhi, Hauz Khas, New Delhi 110016, India}

\author{Kenji Watanabe}
\affiliation{Research Center for Functional Materials, National Institute for Materials Science, 1-1 Namiki, Tsukuba 305-0044, Japan }

\author{Takashi Taniguchi}
\affiliation{ International Center for Materials Nanoarchitectonics, National Institute for Materials Science, 1-1 Namiki, Tsukuba 305-0044, Japan }

\author{Hicham Majjad}
\affiliation{Universit\'e de Strasbourg, CNRS, Institut de Physique et Chimie des Mat\'eriaux de Strasbourg, UMR 7504, F-67000 Strasbourg, France}

\author{Michelangelo Romeo}
\affiliation{Universit\'e de Strasbourg, CNRS, Institut de Physique et Chimie des Mat\'eriaux de Strasbourg, UMR 7504, F-67000 Strasbourg, France}

\author{Arnaud Gloppe}
\affiliation{Universit\'e de Strasbourg, CNRS, Institut de Physique et Chimie des Mat\'eriaux de Strasbourg, UMR 7504, F-67000 Strasbourg, France}

\author{Ludger Wirtz}
\affiliation{Department of Physics and Materials Science, University of Luxembourg, 162a avenue de la Fa\"iencerie, L-1511 Luxembourg, Luxembourg }

\author{St\'ephane Berciaud}
\email{stephane.berciaud@ipcms.unistra.fr}
\affiliation{Universit\'e de Strasbourg, CNRS, Institut de Physique et Chimie des Mat\'eriaux de Strasbourg, UMR 7504, F-67000 Strasbourg, France}


\begin{abstract}

Quantised lattice vibrations (\textit{i.e.}, phonons) in solids are robust and unambiguous fingerprints of crystal structures and of their symmetry properties. In metals and semimetals, strong electron-phonon coupling may lead to so-called Kohn anomalies in the phonon dispersion, providing an image of the Fermi surface in a non-electronic observable. Kohn anomalies become prominent in low-dimensional systems, in particular in graphene, where they appear as sharp kinks in the in-plane optical phonon branches.
However, in spite of intense research efforts on electron-phonon coupling in graphene and related van der Waals heterostructures, little is known regarding the links between the symmetry properties of optical phonons at and near Kohn anomalies and their sensitivity towards the local environment. Here, using inelastic light scattering (Raman) spectroscopy, we investigate a set of custom-designed graphene-based van der Waals heterostructures, wherein dielectric screening is finely controlled at the atomic layer level. We demonstrate experimentally and explain theoretically that, depending exclusively on their symmetry properties, the two main Raman modes of graphene react differently to the surrounding environment. While the Raman-active near-zone-edge optical phonons in graphene undergo changes in their frequencies due to the neighboring dielectric environment, the in-plane, zone-centre optical phonons are symmetry-protected from the influence of the latter. These results shed new light on the unique electron-phonon coupling properties in graphene and related systems and provide invaluable guidelines to characterise dielectric screening in van der Waals heterostructures and moir\'e superlattices.

\end{abstract}


\maketitle


%


\section{Introduction}

Phonons in insulators and semiconductors, such as diamond and silicon, respectively~\cite{Solin1970,Temple73}, are routinely exploited for calibration and sensing purposes and generally considered as weakly sensitive to the local environment. In the particular case of metals, strong coupling between electrons and optical phonons with specific nesting wavevectors can give rise to Kohn anomalies in the phonon dispersion~\cite{Kohn1959}. Kohn anomalies may drive phase transitions such as charge density waves and also provide an image of the Fermi surface~\cite{Baroni2001}. In addition, Kohn anomalies may be very sensitive to screening effects, through the screened electron-phonon coupling. However, the fundamental links between symmetries and Kohn anomalies deserve further attention. These links become prominent and experimentally accessible in reduced dimensions, for instance in low-dimensional carbon-based materials, e.g. carbon nanotubes~\cite{Piscanec2007} and graphene~\cite{Piscanec2004,Lazzeri2008}. Interestingly, the rise of two-dimensional (2D) materials beyond graphene (such as transition metal dichalcogenides (TMD)~\cite{Wang2018} and hexagonal boron nitride (hBN)~\cite{Roy2021}) and of the concept of van der Waals engineering~\cite{Geim2013,Lau2022} make it possible to devise quantum materials and devices, wherein dielectric screening can be finely controlled at the atomic monolayer level~\cite{Raja2017,Lorchat2020}, to uncover and exploit the subtle interplay between electronic and vibrational degrees of freedom.


Here, taking monolayer graphene as a model system, we show that the two Kohn anomalies that occur at the centre ($\mathbf{\Gamma}$) and edges ($\mathbf{K}$ and $\mathbf{K}^{\prime}$ points) of its Brillouin zone, respectively (Fig.~\ref{fig1}a), behave differently when the local dielectric environment is changed, just because the associated phonon modes behave differently under symmetry operations. For this purpose, we investigate, using micro-Raman scattering spectroscopy~\cite{Ferrari2013}, the G-,~2D-~and~2D$^{\prime}$-mode features in monolayer graphene embedded in a carefully chosen variety of environments, evolving from the unscreened, freely suspended case to a van der Waals heterostructure, where graphene is tightly coupled to a TMD monolayer and fully encapsulated in between thin films of hBN. We unravel a progressive upshift of the 2D-mode feature of up to 26~cm$^{-1}$ that we attribute to the screening of the Kohn anomaly that affects non-degenerate zone-edge in-plane transverse optical phonons (TO branch at $\mathbf{K}$ and $\mathbf{K}^{\prime}$, with $A^{\prime}_1$ symmetry). Importantly, we find that screening arises for the most part from the first adjacent layer and that a TMD monolayer screens more efficiently than bulk hBN. Measurements of the dispersion of the 2D-mode frequency reveal that the effect of screening is more pronounced as the wavevector of the 2D-mode phonon approaches $\mathbf{KK}^{\prime}$. Remarkably, we demonstrate that the zone-centre in-plane optical phonons (degenerate LO/TO branch at $\mathbf{\Gamma}$, with $E_{\mathrm{2g}}$ symmetry) giving rise to the Raman G-mode feature are, to experimental accuracy, insensitive to dielectric screening. These contrasting behaviours are explained theoretically by considering the symmetry properties of low-energy electronic and electron-phonon Hamiltonians in graphene and  showing that the zone-centre phonon frequency is protected from changes in the screening by a Ward identity, which leads to a cancellation of many-body renormalization effects in the interatomic force constants. Our results shed light on electron-phonon physics and demonstrates how symmetry is crucial to understand the protection of certain observables, including the frequency of optical phonons. From a practical standpoint, the protection of the G-mode frequency combined with the high sensitivity of the 2D-mode feature with respect to dielectric screening provide an invaluable missing element for advanced characterization of graphene and more broadly of near-field coupled van der Waals materials.

\section{Raman spectroscopy of graphene in various dielectric environments}

\begin{figure*}[t!]
    \begin{center}
    \includegraphics[width=0.92\linewidth]{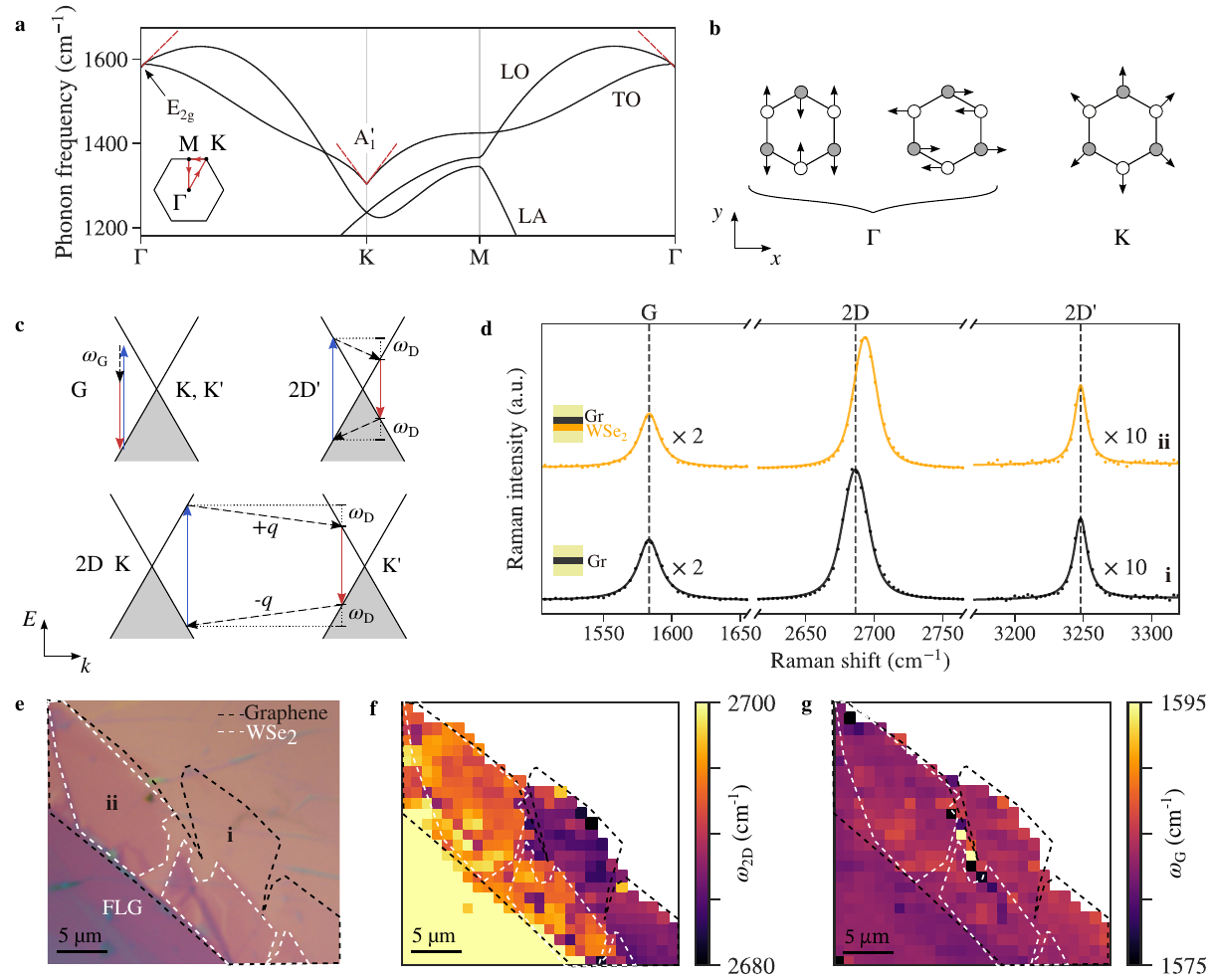}
    \caption{\textbf{Symmetry-dependent screening-induced changes in the Raman spectrum of graphene.} (a) Dispersion of the higher energy phonon branches in graphene along the high-symmetry lines of its first Brillouin zone (see inset), adapted from~\cite{reichardt2018}. The Kohn anomalies at $\mathbf{K}$ and $\mathbf{\Gamma}$ are highlighted with red dashed lines. (b) In-plane atomic displacements associated with the $E_{\mathrm{2g}}$ and $A^{\prime}_1$ phonons. (c) Sketches of the Raman G-, 2D-, and 2D$^{\prime}$- modes processes in the momentum-energy space.  (d) Typical Raman spectra of hBN-capped graphene (black dots) and hBN-capped WSe$_2$/graphene (orange dots). Data are shown together with their fit (solid lines). The layered structures are sketched with the encapsulating hBN films in light yellow. (e) Optical image of the corresponding sample (Sample 1). The contours of the relevant regions of the sample are indicated. FLG denotes a few-layer graphene flake. (f,g) Hyperspectral Raman maps of the 2D-mode and G-mode frequencies in the sample shown in (e), respectively.}
    \label{fig1}
    \end{center}
\end{figure*}

Raman scattering spectroscopy is known as a unique probe of the physical properties of graphene and van der Waals materials~\cite{Ferrari2013,Zhangx2015}, able to yield decisive insights into their nanoscale environment~\cite{Gadelha2021}, symmetry properties~\cite{Chakraborty2012,Froehlicher2015b} and sensitivity towards external perturbations~\cite{Ryu2010}. Monolayer graphene stands out as the first~\cite{Ferrari2006,Graf2007} and most studied 2D material, largely because of the strong coupling between its electronic states and its optical phonon modes. More precisely, the Fermi surface of neutral graphene (i.e., the $\mathbf{K},\:\mathbf{K^{\prime}}$ points) enables Kohn anomalies for phonon wavevectors $q=0$ and $\mathbf{q}=\mathbf{KK}^{\prime}$~, which appear as sharp kinks in the phonon dispersion~\cite{Piscanec2004,Lazzeri2008} (Fig.~\ref{fig1}a,b).%

The Raman spectrum of graphene (Fig.~\ref{fig1}c,d) is dominated by two main features. First, the one-phonon, non-resonant G-mode feature (frequency $\wG$ near 1580~cm$^{-1}$) arising from degenerate LO/TO $E_{\mathrm{2g}}$ phonons at $\mathbf{\Gamma}$~\cite{Basko2009b}. Second, strong electron-phonon coupling favors resonant multiphonon processes~\cite{Maultzsch2004,Basko2008,Venezuela2011}, the most famous of which being the 2D mode, with a pair of near zone-edge ($\mathbf{K}$, $\mathbf{K^{\prime}}$) TO phonons with opposite momenta that warrants conservation of energy and momentum (Fig.~\ref{fig1}c). The 2D-mode feature is widely known as a hallmark of monolayer graphene~\cite{Ferrari2006,Graf2007}; it is also dispersive (frequency $\wDD$ between 2500 and 2700~cm$^{-1}$ for near-IR to near UV incoming photons) and thereby enables probing both the electron and phonon dispersions in various environments~\cite{Mafra2007,Berciaud2013}. A conceptually similar mode involving a pair of LO phonons near $\mathbf{\Gamma}$ gives rise to the fainter 2D$^{\prime}$ mode feature near 3240~cm$^{-1}$ (Ref.~\onlinecite{Basko2008,Venezuela2011}, Fig.~\ref{fig1}c and Appendix~\ref{Section2D'}).

The sensitivity of the G- and 2D-mode features upon strain~\cite{Mohiuddin2009}, doping~\cite{Pisana2007,Yan2007,Das2008,Chen2011,Froehlicher2015} and defects~\cite{Cancado2011} has been extensively documented, whereas studies of dielectric screening effects are scarce~\cite{Forster2013,Berciaud2013,Froehlicher2018}. Here, crucially, in order to isolate the role of dielectric screening, we have focused on samples that have been thoroughly characterised by means of hyperspectral Raman mapping and that display negligible unintentional doping, built-in strain and defect density (Supplemental Material, Sections S1 and S2~\cite{SMnote}).

Figure \ref{fig1}e shows an optical image of a graphene monolayer, partly covered by a tungsten diselenide (WSe$_2$) monolayer and fully encapsulated in hBN (Sample 1, see Appendix~\ref{Methods} for details on the experimental methods and the Supplemental Material~\cite{SMnote}). Typical Raman spectra associated with both regions of the sample (Fig.~\ref{fig1}d) evidence a clear shift of $\wDD$, while at the same time, the G-mode and the 2D$^{\prime}$-mode features remain unaffected. An hyperspectral map of $\wDD$ is presented in Fig.~\ref{fig1}f and readily confirms that hBN-capped graphene/WSe$_2$ exhibits higher values of $\wDD$  than hBN-capped graphene. In order to quantitatively characterize the role of the environment, we spatially average $\wDD$ and $\wG$ over an area of a few tens of µm$^2$ in regions (i) and (ii) in Fig.~\ref{fig1}e. We find a spatially averaged upshift of $\wDD$ of $7.3 \pm 3 ~\wn$, when comparing hBN-capped WSe$_2$/graphene to hBN-capped graphene. Noteworthy,  the spatial averages of ($\wG=1583.7\pm 0.8~\wn$ and $\wG=1583.5\pm 0.5~\wn$ in hBN-capped graphene and graphene/WSe$_2$, respectively) are nearly identical. These contrasting behaviours directly exclude a simple interpretation in terms of distinct levels of strain and/or doping in both regions, as one would expect a G-mode upshift of $\approx 3.3~\wn$ in the case of biaxial compressive strain~\cite{Metten2014,Androulidakis2015} and a much larger upshift shift in excess of $10~\wn$ in the case of electron or hole doping~\cite{Lee2012,Froehlicher2015}. A more plausible explanation is that the presence of one single layer of TMD screens the Kohn anomaly at $\mathbf{K}$ and $\mathbf{K^{\prime}}$~\cite{Endlich2013}, leading to a sizeable upshift of $\wDD$~\cite{Forster2013,Froehlicher2018}, as schematised in Fig.~\ref{figDisp}.

\section{Impact of dielectric screening near the Kohn anomaly at K}

\begin{figure*}[t!]
    \begin{center}
    \includegraphics[width=0.92\linewidth]{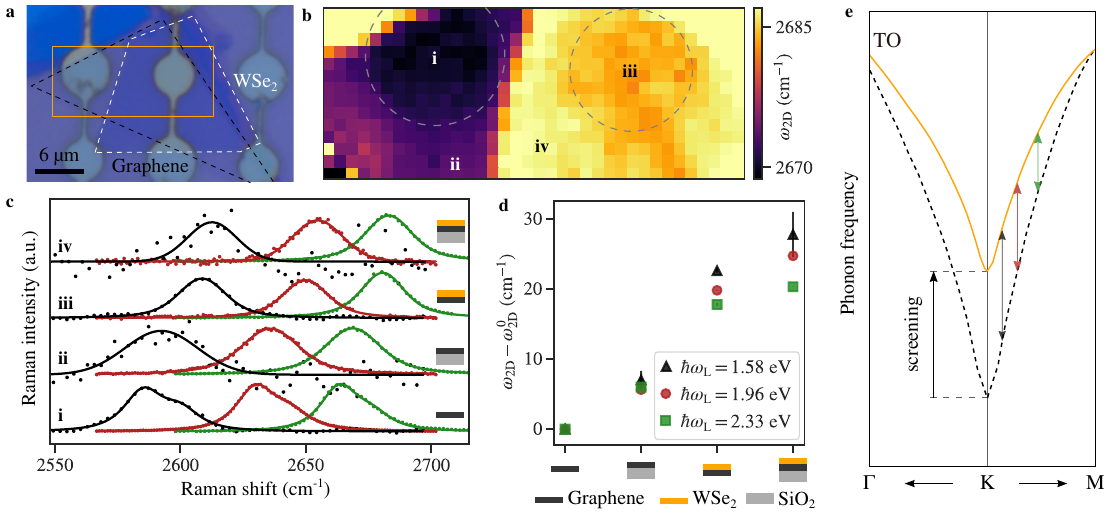}
    \caption{\textbf{Probing dielectric screening of the Kohn anomaly at the K and K$^{\prime}$ points.} (a) Optical image of a graphene/WSe$_2$ heterostructure deposited on an Si/SiO$_2$ substrate with pre-patterned holes (Sample 2). The white dashed contour indicates the WSe$_2$ monolayer. The scalebar is $6~\mu\mathrm{m}$. Four regions of interest corresponding to freely suspended graphene (i), graphene on SiO$_2$ (ii), suspended graphene/WSe$_2$ (iii), and SiO$_2$-supported graphene/WSe$_2$ (iv) are identified.  (b) Hyperspectral Raman map of the graphene 2D-mode frequency $\omega_{\mathrm{2D}}$. The mapped area corresponds to the dark yellow rectangle in (a). (c) Raman 2D-mode spectra of the sample shown in (a) for different laser energies: 1.58~eV (black), 1.96~eV (red) and 2.33~eV (green). Data are shown together with their fit. (d) 2D-mode frequency $\omega_{\mathrm{2D}}$ in the different regions of the sample as a function of the (symbol-coded) laser photon energy. Data are shown relative to $\omega_{\mathrm{2D}}^0$, the value recorded on suspended graphene. (e) Sketch of the TO phonon dispersion in the vicinity of the $\mathbf{K}$ point for unscreened (dashed black line) and screened graphene (solid orange line). The vertical arrows indicate the expected screening-induced phonon upshift for phonon wavevectors corresponding to increasing laser photon energy. Dielectric screening flattens the Kohn anomaly (kink at $\mathbf{K}$).}
    \label{figDisp}
    \end{center}
\end{figure*}

To get deeper insights into the influence of the dielectric environment, we now consider Sample 2, made of a monolayer graphene deposited onto a  SiO$_2$ substrate patterned with microscopic circular pits and partly covered by a WSe$_2$ monolayer. Fig.~\ref{figDisp}a shows an optical image of this sample, where one can identify four regions of interest, namely, freely suspended graphene (i), graphene on SiO$_2$ (ii), suspended graphene/WSe$_2$ (iii), and SiO$_2$-supported graphene/WSe$_2$ (iv). The interest of such a sample is threefold. First, it makes it possible to study suspended graphene as a pristine, minimally screened sample~\cite{Berciaud2009,Faugeras2015}. Second, the screening induced by one single TMD monolayer can be easily identified in suspended WSe$_2$/graphene. Third, by using SiO$_2$ as a bulk substrate with similar dielectric constant as bulk hBN~\cite{Geick1966,Volksen2010}, we will be able to determine to which extent surface roughness affects the observed 2D-mode upshift introduced above.

Figure~\ref{figDisp}b shows an hyperspectral map of $\wDD$ (under laser excitation at 2.33~eV), that, again, evidences a clear progressive upshift when moving from  suspended graphene to SiO$_2$-supported graphene/WSe$_2$, i.e., from region (i) to (iv) (Fig.~\ref{figDisp}b,c). In stark contrast, the spatially averaged $\wG$ are nearly identical on all four regions, spread only within a 2~cm$^{-1}$ window (see Supplemental Material, Section S2~\cite{SMnote}), in keeping with the trend observed in Fig.~\ref{fig1}.

Remarkably, by investigating the dispersion of $\wDD$ using laser excitation at 2.33~eV, 1.96~eV and 1.58~eV in the four regions identified above (Fig.~\ref{figDisp}c), we observe that the screening-induced 2D-mode upshift augments as the laser photon energy decreases. In Fig.~\ref{figDisp}d, we plot, in each region, the laser photon energy dependent upshift of $\wDD$, with color-coded symbols. Under laser excitation at 2.33~eV (resp. 1.58~eV),  $\wDD$ upshifts by up to $20~\wn$ (resp. $27~\wn$) when moving from region (i) to (iv). These observations are consistent with the screening of the Kohn anomaly illustrated in Fig.~\ref{figDisp}e. Indeed, as the laser photon energy increases, the momentum  of the phonons involved in the 2D-mode process (taken relative to $\mathbf{K}$) also increases and the screening-induced upshift of phonon frequency fades away.

Thus far, the 2D-mode upshift has been interpreted as being the result of a change of the phonon dispersion when the screening is varied~\cite{Endlich2013,Forster2013}. In general though, the 2D~mode process strongly relies on the resonant interplay between the electronic and phonon dispersions (in a simple one-dimensional description with linear bands and no trigonal warping effects, one gets $\wDD=2 \omega_{\mathbf{K}}+2\frac{v_{\mathrm{TO}}}{v_{\mathrm{F}}}\omega_{\mathrm{L}}$, with $\omega_{\mathbf{K}}$, $v_{\mathrm{TO}}$, $v_{\mathrm{F}}$ and $\omega_{\mathrm{L}}$, the TO phonon frequency at $\mathbf{K}$, the phonon and electron band velocities, and the laser photon frequency, respectively). As a result, the observed 2D-mode upshift may also be affected by the partial compensation between the screening-induced reduction of the phonon and electron band velocities~\cite{Mafra2007,Berciaud2013,Maultzsch2004}. Nevertheless, the results in Fig.~\ref{figDisp}d, which reveal the dispersive character of the \textit{relative} dielectric shift, also indicate that in the energy range probed in our study, this dispersive effect appears only as a correction to main effect that is a rigid upshift of $\omega_{\mathbf{K}}$ when the dielectric screening increases. Further measurements at lower $\omega_{\mathrm{L}}$ would allow probing 2D-mode phonons closer to the $\mathbf{K}$ point~\cite{Venanzi2023} and thereby get more insights into the upshift of $\wDD$.

\section{Evidence for symmetry-dependent dielectric screening of optical phonons}


\begin{figure*}[t!]
    \begin{center}
    \includegraphics[width=0.65\linewidth]{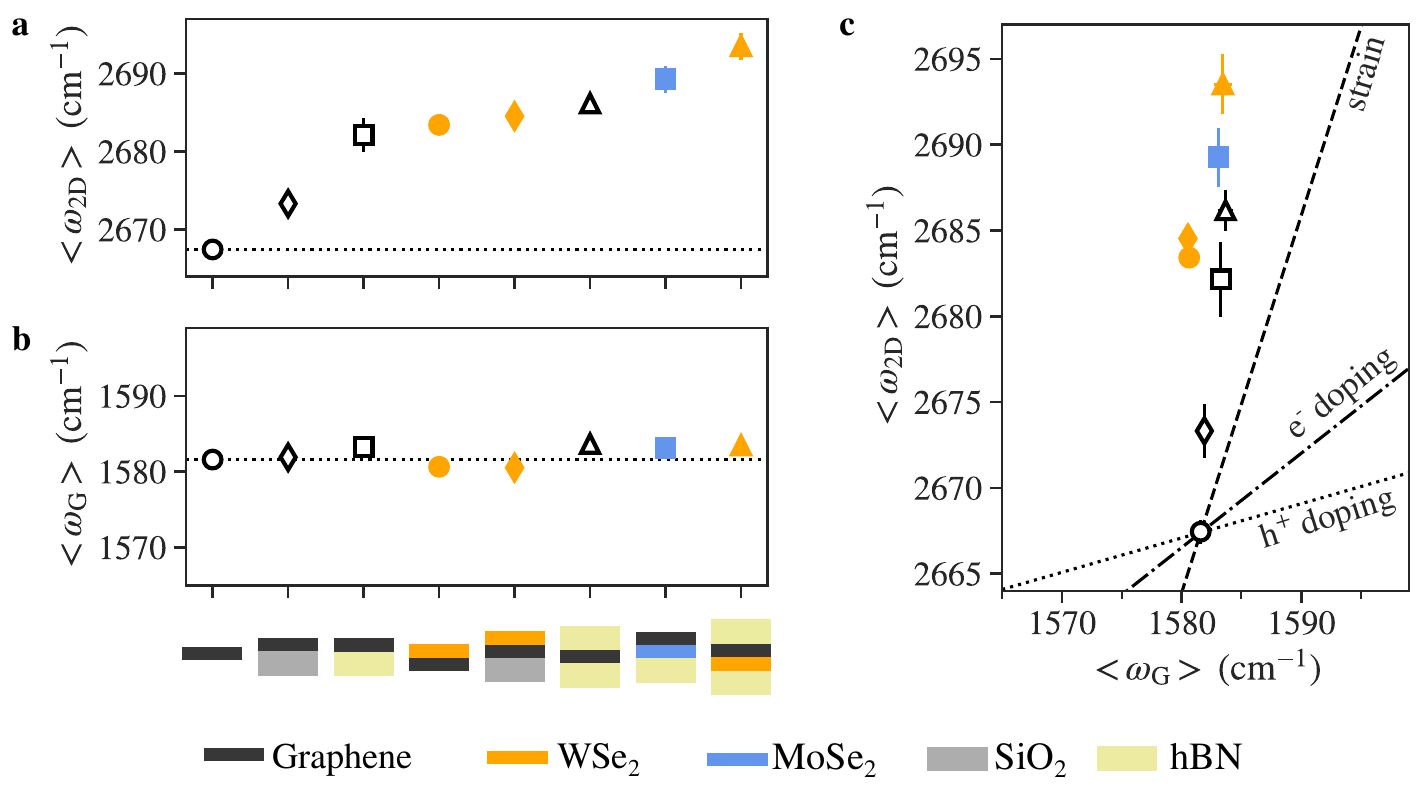}
    \caption{\textbf{Upshift of the 2D-mode frequency and invariance of the G-mode frequency in various dielectric environments. } Spatially averaged 2D-mode (a) and G-mode (b) frequencies measured on a variety of graphene-based van der Waals heterostructures (see sketches below the graphs; the bare graphene monolayer and graphene/WSe$_2$ heterostructure -first and fourth samples- correspond to the freely suspended systems discussed in Fig.~\ref{figDisp}). Samples are ordered by increasing expected dielectric screening experienced by graphene. (c) Correlation between the spatially averaged 2D- and G-mode frequencies shown in (a) and (b). For reference, the short dashed, dot-dashed and dotted lines correspond to the expected correlations in the presence of biaxial strain, hole and electron doping, respectively. The reference point for undoped and unstrained graphene is taken for the suspended graphene monolayer. All measurements were performed under laser excitation at 2.33~eV.}
    \label{figRaja}
    \end{center}
\end{figure*}

The spatially averaged values of $\wDD$ and $\wG$ obtained from Sample 1 (Fig.~\ref{fig1}) and Sample 2 (Fig.~\ref{figDisp}) as well as on another sample (BN/MoSe$_2$/graphene, Sample 3 introduced in the Supplemental Material, Section S2~\cite{SMnote}, see also  Ref.~\onlinecite{parralopez2021}) are gathered in Fig.~\ref{figRaja}a and b, respectively. The areas of interest are ordered by increasing expected dielectric screening. Errorbars correspond to the standard deviation of the spatially-averaged values of $\wDD$ and $\wG$. 

While the upshift of $\wDD$ due to the bulk SiO$_2$ substrate itself is only of $5.9 \pm 2.3~\wn$, we find much larger upshifts of $14.8\pm2.8~\wn$ and $16.0\pm1.2~\wn$ when SiO$_2$ is replaced by van der Waals materials, namely bulk hBN and monolayer WSe$_2$, respectively. In spite of their similar dielectric constants~\cite{Geick1966,Volksen2010}, bulk hBN leads to a much larger shift of $\wDD$ that bulk SiO$_2$.  As SiO$_2$ is rougher than a layered material such as hBN, this observation suggests that graphene remains partly decoupled from SiO$_2$, whereas hBN, as an atomically flat substrate, warrants optimal coupling and drastically reduces surface roughness and disorder in graphene~\cite{Xue2011}. This hypothesis is corroborated by the particularly narrow 2D-mode full width at half maximum of only 16~cm$^{-1}$ observed in hBN-capped graphene and hBN-capped graphene/WSe$_2$ (Appendix~\ref{section_FWHM} and Ref.~\onlinecite{Neumann2015,Neumann2016}) and by a comparative study of SiO$_2$/graphene/TMD versus SiO$_2$/TMD/graphene heterostructures (Appendix~\ref{Sec_tuning_roughness}).
Noteworthy, the screening-induced upshift of $\wDD$ is slightly larger in suspended graphene/monolayer WSe$_2$ than in graphene deposited on bulk hBN (Fig.~\ref{figRaja}a and Supplemental Material, Section S2~\cite{SMnote}), illustrating the remarkably strong interaction between a TMD monolayer and graphene, as suggested otherwise by recent electron transport measurements~\cite{Banszerus2019}.  Altogether, this set of results demonstrates that the screening of the Kohn anomaly at $\mathbf{K}$ is intimately linked to the interfacial coupling between graphene and the nearest atomic layer. This conclusion is bolstered by our demonstration that a TMD bilayer does only lead to a marginal increase of $\wDD$ compared to a TMD monolayer (Appendix~\ref{Sec_tuning_bilayer}). 

Let us stress that the screening of optical phonons also stands out as an extremely short-range effect (typically, the sub-nanometre van der Waals gap between coupled 2D layers), whereas it was shown that excitons could undergo the effects of dielectric screening at larger distances of several nanometres (compare Fig.~1c and 1d Ref. \onlinecite{Froehlicher2018} and see also Ref.~\onlinecite{Tebbe2023}). We attribute this difference to the fact that for the excitons, the excited electron interacts both with the electrostatic potential of the hole as well as with the potential of the screening cloud originating from all over space, and in general from several neighboring layers (Appendix~\ref{Sec_tuning_PhononvsExciton} for details).

We now analyze the contrasting behaviour introduced in Fig.~\ref{fig1} between Raman modes involving optical phonons at or near $\mathbf{\Gamma}$ (G- and 2D$^{\prime}$-modes, respectively) and near $\mathbf{K},~\mathbf{K}^{\prime}$ (2D-mode). As demonstrated in Fig.~\ref{figRaja}a,b, even in very different dielectric environments, $\wG$ remains constant at $1582\pm2~\wn$ (fluctuations that can be ascribed to built-in strain and unintentional doping, see Supplemental Material, Section S2~\cite{SMnote}), whereas $\wDD$ upshifts by more than $20~\wn$. The invariance of $\wG$ is further evidenced on the  $\wDD$ versus $\wG$ correlation plot displayed in Fig.~\ref{figRaja}c. A linear fit to the plotted data yields a slope $\partial\wDD/\partial\wG$ above 10, that largely exceeds the expected slopes in the presence of strain~\cite{Metten2014,Androulidakis2015} or doping~\cite{Lee2012,Froehlicher2015}. This behaviour is \textit{a priori} counter intuitive as a Kohn anomaly is also present at $\mathbf{\Gamma}$, suggesting that a sizeable upshift of $\wG$ (and $\omega_{\mathrm{2D^{\prime}}}$)  may be in order when graphene is screened by a partner material.



\section{Theoretical argument}

The insensitivity of the $\mathbf{\Gamma}$-point phonon frequency to changes in the dielectric environment can be understood from a symmetry point of view.
For this, we focus on the contribution of the low-energy $\pi$-band excitations to the part of the phonon frequency that is sensitive to screening~\cite{reichardt2018,Giustino2017}:
\begin{widetext}
\begin{equation}
    \omega^2_{\lambda}|_{\pi\text{-bands}} = \frac{2}{M_{\mathrm{C}}} \int_{\mathrm{BZ}} \frac{\mathrm{d}^2 k}{A_{\mathrm{BZ}}} \sum_{\mathrm{spin}}
        \frac{ \left(g_{\mathrm{bare}}^{\lambda}\right)^*_{\mathbf{k},\pi^*,\pi}  \left( g_{\mathrm{screened}}^{\lambda} \right)_{\mathbf{k},\pi^*,\pi} } { \varepsilon_{\mathbf{k},\pi^*} - \varepsilon_{\mathbf{k},\pi} }.
 \label{eqD}
\end{equation}
\end{widetext}
Here, $\lambda=x,y$ is a label for the phonon polarization, $g^{\lambda}_{\mathrm{bare}}$ and $g^{\lambda}_{\mathrm{screened}}$ are the unscreened and screened electron-phonon coupling matrix elements, respectively, $\varepsilon_{\mathbf{k},\pi^{(*)}}$ denote the band energies, $M_{\mathrm{C}}$ is the mass of a carbon atom, and the integral runs over the first Brillouin zone, whose area is denoted as $A_{\mathrm{BZ}}$.
While the exact definitions of the electron-phonon matrix elements are given in Appendix~\ref{SectionTheory}, we note that the screened electron-phonon coupling and the $\pi$-band transition energies are not independent of one another.
The lattice symmetry of the graphene lattice leads to the effective $\pi$-band Hamiltonian being of the Dirac-Weyl type, while the $E_{2\mathrm{g}}$ symmetry of the phonon implies that it transforms like a vector under rotations, in contrast to the non-degenerate phonon at $K$, which transforms like a pseudo-scalar under rotations.
This "vector-like" character of the phonon mode in combination with the Dirac-Weyl nature of the lead to the effective electron-phonon interaction Hamiltonian being of the same form as the electron-photon Hamiltonian in Quantum Electrodynamics (QED), in which the vector-like phonon is coupled to an electronic current (Appendix~\ref{SectionTheory}).
Crucially, for Dirac electrons, this current is conserved, which on the quantum mechanical level implies that any correlation functions involving this current obey the Schwinger-Dyson equations (see e.g., Ref.~\cite{peskin1995}).
Noting that for the $\pi$-band electrons, the screened electron-phonon coupling can be expressed in terms of a correlation function involving the conserved current, an application of the Schwinger-Dyson equation leads to a Ward identity~\cite{peskin1995,srednicki2007} for the screened electron-phonon coupling.
For electrons near the $K$ valley, it reads:
\begin{equation}
    \left( g_{\mathrm{screened}}^{\lambda} \right)_{\mathbf{k},\pi^*,\pi} \propto v_{\mathrm{F}} \, \left[ \chi^\dagger_{\mathbf{k}\pi^*} \cdot \boldsymbol{\sigma} \cdot \chi_{\mathbf{k},\pi} \right] \wedge \mathbf{e}_{\lambda},
\end{equation}
where $\mathbf{e}_{\lambda=x,y}$ is the effective polarization vector of the phonon, $\boldsymbol{\sigma}$ is the vector of Pauli matrices, and $\chi$ denotes a two-component Weyl spinor in sublattice space, and $\mathbf{a} \wedge \mathbf{b} = a_x b_y - a_y b_x$ is the two-dimensional anti-symmetric product.
Most importantly, the only screening-dependent, \textit{i.e.}, substrate-sensitive, quantity here is the effective Fermi velocity $v_{\mathrm{F}}$.
Notably, the bare electron-phonon coupling is naturally independent of screening and $v_{\mathrm{F}}$, while the electronic transitions energies in the denominator of Eq.~\eqref{eqD}, are proportional to the very same $v_{\mathrm{F}}$: $\varepsilon_{\mathbf{k},\pi^*} - \varepsilon_{\mathbf{k},\pi} \propto v_{\mathrm{F}} |\mathbf{k}|$.
As such, the screened electron-phonon coupling for the $\mathbf{\Gamma}$-point optical phonon as well as the electron $\pi$-band slope and thus all screening effects from the dielectric environment in Eq.~\eqref{eqD} compensate each other.
This exact cancellation for the low-energy $\pi$-band electrons is a result purely of the lattice symmetry, leading to the Dirac-like effective electronic and electron-phonon Hamiltonians, and of current conservation, which is tied to a global $U(1)$ gauge symmetry. It is in excellent agreement with our experimental findings, which show no noticeable change of $\omega_G$ when the dielectric environment is changed. Note that for acoustic phonons, a stronger version of the gauge symmetry exists, which slightly simplifies the discussion in that case~\cite{Basko2008b,Sohier2014} (Appendix~\ref{SectionTheory}).

Consistently, the 2D${^\prime}$ mode, involving a pair of LO phonons near $\boldsymbol{\Gamma}$, albeit being of similar nature as the 2D mode~\cite{Basko2008,Venezuela2011}, is, to experimental accuracy, unaffected by changes in the dielectric environment, due to the protection of the LO phonon frequency at $\boldsymbol{\Gamma}$ (Appendix~\ref{Section2D'}). Since the phonons involved in the 2D${^\prime}$ mode are not strictly at $\boldsymbol{\Gamma}$, this result illustrates the robustness of the symmetry argument introduced above.

We want to stress once more that, while a more complicated and generalized form of the Ward identity can always be derived for the screened electron-phonon vertex in any system, its usefulness in the case of graphene is due to the symmetry of the graphene lattice, which ensures that the electron-phonon interaction for the zero-momentum optical in-plane optical phonons takes on the specific form of an electronic current coupled to a phonon field that transforms like a vector under rotations. This fact makes the Ward identity useful to derive insights into the influence of the dielectric environment on the G-mode phonon. Conversely, for non-degenerate phonons, which necessarily do not transform like a vector under rotations and hence cannot couple to the electronic current, the more general form of the Ward identity, cannot be used to derive any straight insight. In particular, we thus expect that the Raman G-mode frequency will no longer be protected if it becomes non-degenerate, as is, for instance, the case in carbon nanotubes and in uniaxially strained graphene. The latter point also specifically applies to the frequency of the $A^{\prime}_1$ phonon at $\mathbf{K}$ in graphene, which, while also influenced by a Kohn anomaly, is not protected from changes in the screening environment. As a result, there is no symmetry-based reasoning that protects the 2D-mode frequency.

The symmetry-based arguments discussed above could in principle also affect the Raman scattering intensities. Experimentally, we observe that the integrated intensity ratio between the 2D- and G-mode features ($I_{\rm{2D}}/I_{\rm{G}}$) is indeed reduced when comparing freestanding graphene with a TMD/graphene heterostructure (see Supplemental Material, Section S3 for a detailed analysis~\cite{SMnote}). Although computing the Raman scattering intensities of the G- and 2D-modes as a function of screening remains a challenge, the aforementioned observation is in qualitative agreement with the stronger sensitivity of the $A^{\prime}_1$ phonons at $\mathbf{K}$ (and of its coupling to electrons) to dielectric screening.


\section{Conclusion and outlook}

We have demonstrated that the Raman-active optical phonons in graphene are an invaluable gauge of dielectric screening and, importantly of the microscopic details of interlayer coupling within van der Waals heterostructures. The presence of a surrounding medium screens the Kohn anomaly near the $\mathbf{K}$ point, yielding a sizeable upshift of the Raman 2D-mode feature by up to $1\ \%$ when comparing a suspended graphene monolayer with a monolayer graphene/monolayer TMD heterostructure encapsulated in hexagonal boron nitride. Conversely, $E_{\mathrm{2g}}$ phonons, involved in the Raman G-mode and 2D$^{\prime}$-mode process are symmetry-protected from screening effects such that their frequency is invariant in all the experimental configurations examined here.

As an outlook, we believe that the impact of our work can be threefold.

First, the theoretical aspects of our work sheds new light on the deep connections between crystal symmetries, the resulting electronic and vibrational properties, and quantum mechanical conservation laws, which are relevant well-beyond graphene science. In particular, our work provides a general definition of the exact electron-phonon coupling in terms of electronic correlation functions. This is a novel viewpoint on electron-phonon physics, beyond simple models or sophisticated first principles simulations, that will significantly advance ongoing debates on the exact description of the electron-phonon interaction in quantum materials. 

Second, in recent years the correlation between the frequencies of the 2D- and G-mode features has enabled an ever more advanced characterization of graphene-based samples. The comprehensive study undertaken here demonstrates that the observed variability in the 2D-mode frequency can also stem for a large part from changes in the dielectric environment. Thus, using the well-documented sensitivity of the G- and 2D-modes features to strain and doping~\cite{Lee2012a,Metten2014,Froehlicher2015,Metten2017,Neumann2015}, one may straightforwardly isolate the effect of the dielectric environment, which manifests itself as a clear rigid shift of the 2D-mode frequency only. An illustration of such a multiparameter analysis is presented in the Supplemental Material, Section S2.1~\cite{SMnote}. Going further, our results provide invaluable guidelines to decipher hyperspectral Raman maps~\cite{Froehlicher2018}, to better understand proximity effects at van der Waals heterointerfaces~\cite{Lorchat2020} and more broadly to develop advanced Raman scattering-based methods in moir\'e superlattices~\cite{Gadelha2021}.

Finally, our work reveals that the impact of dielectric screening on optical phonons in graphene is chiefly determined by the microscopic details of the interfacial coupling and is extremely short-range, illustrating once again the unique physical properties of van der Waals materials. The extreme sensitivity of the Raman features of graphene towards dielectric screening could be used to develop two-dimensional sensors, able to quantitatively probe interfacial coupling in van der Waals materials with sub-nanometre out-of-plane accuracy, as well as built-in high-pressure gauges~\cite{PiementaMartins2023} as an alternative to ruby~\cite{Chijioke2005}.

\begin{acknowledgements}

We thank T. Sohier for fruitful discussions. We are grateful to the StNano clean room staff for technical support. We acknowledge financial support from the Agence Nationale de la Recherche (under grants ATOEMS ANR-20-CE24-0010 and VANDAMME ANR-21-CE09-0022). This work of the Interdisciplinary Thematic Institute QMat, as part of the ITI 2021 2028 program of the University of Strasbourg, CNRS and Inserm, was supported by IdEx Unistra (ANR 10 IDEX 0002), and by SFRI STRAT'US project (ANR 20 SFRI 0012) and EUR QMAT ANR-17-EURE-0024 under the framework of the French Investments for the Future Program.  S.B., A.R.M., and A.S.  acknowledge support from the Indo-French Centre for the Promotion of Advanced Research (CEFIPRA). S.B. acknowledges support from the Institut Universitaire de France (IUF). K. W. and T. T. acknowledge support from the Elemental Strategy Initiative conducted by the MEXT, Japan, and the CREST (JPMJCR15F3), JST. S.R. acknowledges Fonds National de la Recherche (FNR) Luxembourg project “RESRAMAN” (Grant No. C20/MS/14802965) and L.W. acknowledges support through the FNR project INTER/19/ANR/13376969/ACCEPT.

\end{acknowledgements}



\appendix

\begin{figure*}[t!]
    \begin{center}
    \includegraphics[width=0.75\linewidth]{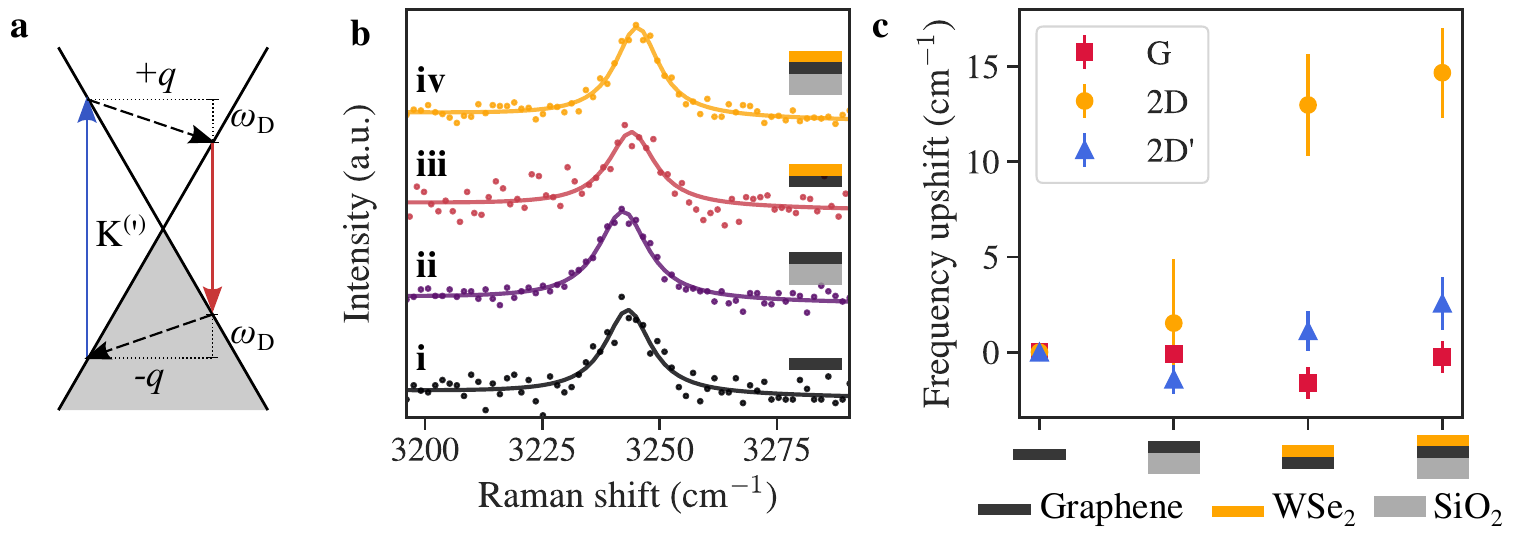}
    \caption{\textbf{Invariance of the 2D$^{\prime}$-mode frequency.} (a) Sketch of the Raman 2D$^{\prime}$-mode process. (b) Spectra of the 2D$^{\prime}$-mode feature in the different configurations considered. The sample and its areas of interest are introduced in Fig.~\ref{figDisp} of main text. (c) Averaged values of 2D$^{\prime}$-, 2D- and G-mode frequencies in each dielectric environment, respectively.}
    \label{fig_2D'}
    \end{center}
\end{figure*}

\section{Experimental methods}
\label{Methods}
All samples were made from mechanically exfoliated bulk crystals of graphite, TMDs (here WSe$_2$, MoSe$_2$ and MoS$_2$) and hBN. The van der Waals heterostructures were deposited onto Si/SiO$_2$ substrates using dry transfer methods, either using directly a PDMS poly(dimethylsiloxane) stamp~\cite{Castellanos-Gomez2014} or through the "pick-up and lift" method, using an additional PC (polycarbonate) layer that is later dissolved in chloroform~\cite{Wang2013,Zomer2014}. Suspended structures were prepared by depositing graphene and WSe$_2$ layers onto Si/SiO$_2$ substrates patterned by $\approx 5~\mu \rm  m$ in diameter pits fabricated using photolithography followed by reactive ion etching~\cite{Berciaud2009}. Raman and photoluminescence (PL) measurements were performed in ambient conditions using a home-built scanning confocal setup as described elsewhere~\cite{Metten2014,Froehlicher2018}, using single mode, linearly polarised lasers emitting at 2.33~eV, 1.96~eV and 1.58~eV. The laser intensity was maintained low enough (i.e., well below $1~\rm{m W}/\mu\rm{m}^2$ for Raman measurements, with even lower intensities for PL measurements) to minimize heating and, importantly, photoinduced doping effects~\cite{Froehlicher2018,parralopez2021}. No polarization analysis was performed on the Raman scattered and PL photons. All samples discussed here have been characterised by means of hyperspectral mapping (Supplemental Material, Sections S1 and S2 ~\cite{SMnote}), allowing unambiguous identification of graphene~\cite{Ferrari2006,Graf2007} and TMD monolayers~\cite{Mak2010}, respectively. PL maps allow to identify coupled and decoupled  TMD/graphene regions based on the strong PL quenching due to non-radiative transfer of TMD excitons to graphene~\cite{Froehlicher2018}. From the spatially-resolved Raman measurements, we can map the built-in strain and non-intentional doping (Supplemental Material, Section S2~\cite{SMnote} and Ref.~\onlinecite{Lee2012,Metten2013}). All samples discussed in this manuscript display small built-in strain of at most a few $10^{-2}\,\%$, while unintentional doping remains below $10^{-12}~\rm {cm}^{-2}$. In these conditions the  shifts of $\wG$ and $\wDD$ due to built-in strain and unintentional doping are negligible compared to the dominant effect of dielectric screening (Fig.~\ref{figRaja}a) and are responsible for the small scatter in the spatially averaged $\wG$ (Fig.~~\ref{figRaja}b). 


\section{Impact of dielectric screening on the Raman 2D' mode}
\label{Section2D'}

The 2D$^{\prime}$ mode is a resonant process that involves two LO phonons with opposite momenta near the $\mathbf{\Gamma}$ point and appears near 3240~$\rm{cm^{-1}}$. It is the \textit{intra}-valley analogue of the 2D mode~\cite{Basko2008,Venezuela2011}. The fully resonant part of the Raman 2D$^{\prime}$ mode process is sketched in Fig.~\ref{fig_2D'}a.  Since the electron-phonon coupling at $\mathbf{\Gamma}$ is smaller than at $\mathbf{K}$, the 2D$^{\prime}$-mode feature is less intense than the 2D-mode feature~\cite{Piscanec2004,Basko2008b}.

In this Appendix, we investigate the evolution of the 2D$^{\prime}$ mode frequency $\omega_{\mathrm{2D^{\prime}}}$ in various dielectric environments to further confirm the behaviour evidenced in Fig.~1d of the main text (Sample 1). For this purpose, we consider the same sample and areas of interest as in Fig.~~\ref{figDisp} of the main text (see also Fig.~S2d-f~\cite{SMnote}), namely suspended graphene (i), SiO$_2$-supported graphene (ii), suspended graphene/WSe$_2$ (iii) and SiO$_2$-supported graphene/WSe$_2$ (iv). 

In the most screened case (i.e., in this sample, SiO$_2$-supported graphene/WSe$_2$), we measure an almost negligible spatially averaged blueshift $\Delta \omega_{\mathrm{2D^{\prime}}} = 2.6 \pm 1.4$ cm$^{-1}$ with respect to the unscreened case of suspended graphene, while, within experimental accuracy, we do not resolve any sizeable shift when comparing the other remaining cases to suspended graphene. We can thus consider that $\Delta \omega_{\mathrm{2D^{\prime}}}$ is negligibly small and in any event more than one order of magnitude smaller than the 2D-mode upshift $\Delta \omega_{\mathrm{2D}}$ presented in Fig.~2. These results are in agreement with the theoretical discussion on the symmetry protection of the optical phonon frequency at $\Gamma$ presented in the main text and in Section~\ref{SectionTheory} below and further demonstrate that the screening of the Kohn anomaly that occurs for zone-edge TO phonons at $\mathbf{K}$ can be used as a sensitive probe of dielectric screening.

\section{2D-mode linewidth}
\label{section_FWHM}

\begin{figure}[h!]
    \begin{center}
    \includegraphics[width=0.9\linewidth]{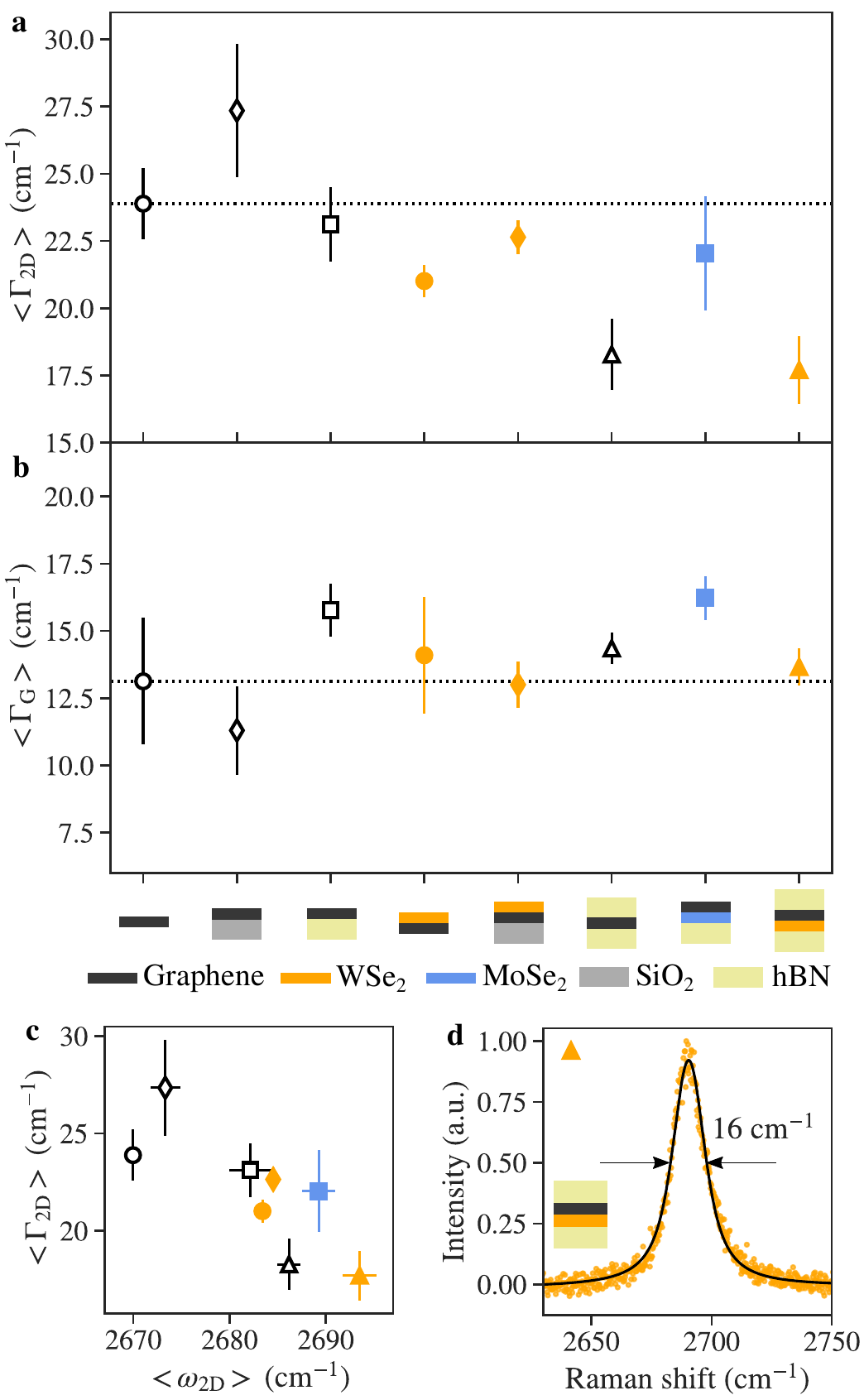}
    \caption{\textbf{Analysis of the 2D-mode linewidth.} (a) Spatially averaged FWHM of the Raman 2D mode of graphene depending of its dielectric environment, ordered by increasing expected dielectric screening. (b) Spatially averaged FWHM of the Raman G mode of graphene depending of its dielectric environment, ordered by increasing expected dielectric screening. (c) Observed correlation between $\langle \GammaDD \rangle$ and $\langle \wDD \rangle$. (d) Observation of a narrow 2D-mode feature in hBN-capped graphene/WSe$_2$. The solid line is a Lorentzian fit.}
    \label{fig_FWHM}
    \end{center}
\end{figure}

The FWHM of the 2D-mode feature of graphene $\GammaDD$ can provide additional information about the dielectric environment of the graphene monolayer. Fig.~\ref{fig_FWHM}a shows $\langle \GammaDD \rangle$ the spatially averaged FWHM of the 2D-mode of graphene depending of its environment, ordered by increasing expected dielectric screening. A sharpening of the 2D mode can be observed as screening increases. In particular, hBN-capping gives greatly reduced linewidths of the 2D mode down to $16 \pm 0.5$ $\wn$ for the sharpest spectra (see Fig.~\ref{fig_FWHM}d). This sharpening of the 2D-mode feature points towards an increased homogeneity of the dielectric environment of graphene, when it is tightly coupled to a layered material~\cite{Neumann2015,Neumann2016}. Interestingly, this reduction of $\GammaDD$ is correlated to the blueshift of $\wDD$, as visible in Fig.~\ref{fig_FWHM}c, highlighting the impact of the interface quality on the overall dielectric environment of graphene. These observations are consistent with our discussion in the main text on the impact of surface roughness on the screening-induced blueshift of the 2D-mode feature, as further discussed in the next section.

\section{Fine tuning of the dielectric environment}
\label{Sec_tuning}
\subsection{Monolayer versus bilayer TMD}
\label{Sec_tuning_bilayer}

To further explore the impact of interfacial coupling on the Raman 2D-mode feature, we compare different areas of another graphene/WSe$_2$ sample (Sample 4, not discussed in Fig.~\ref{figRaja}) with various degrees of coupling between graphene WSe$_2$. 

\begin{figure}[h!]
    \begin{center}
    \includegraphics[width=1\linewidth]{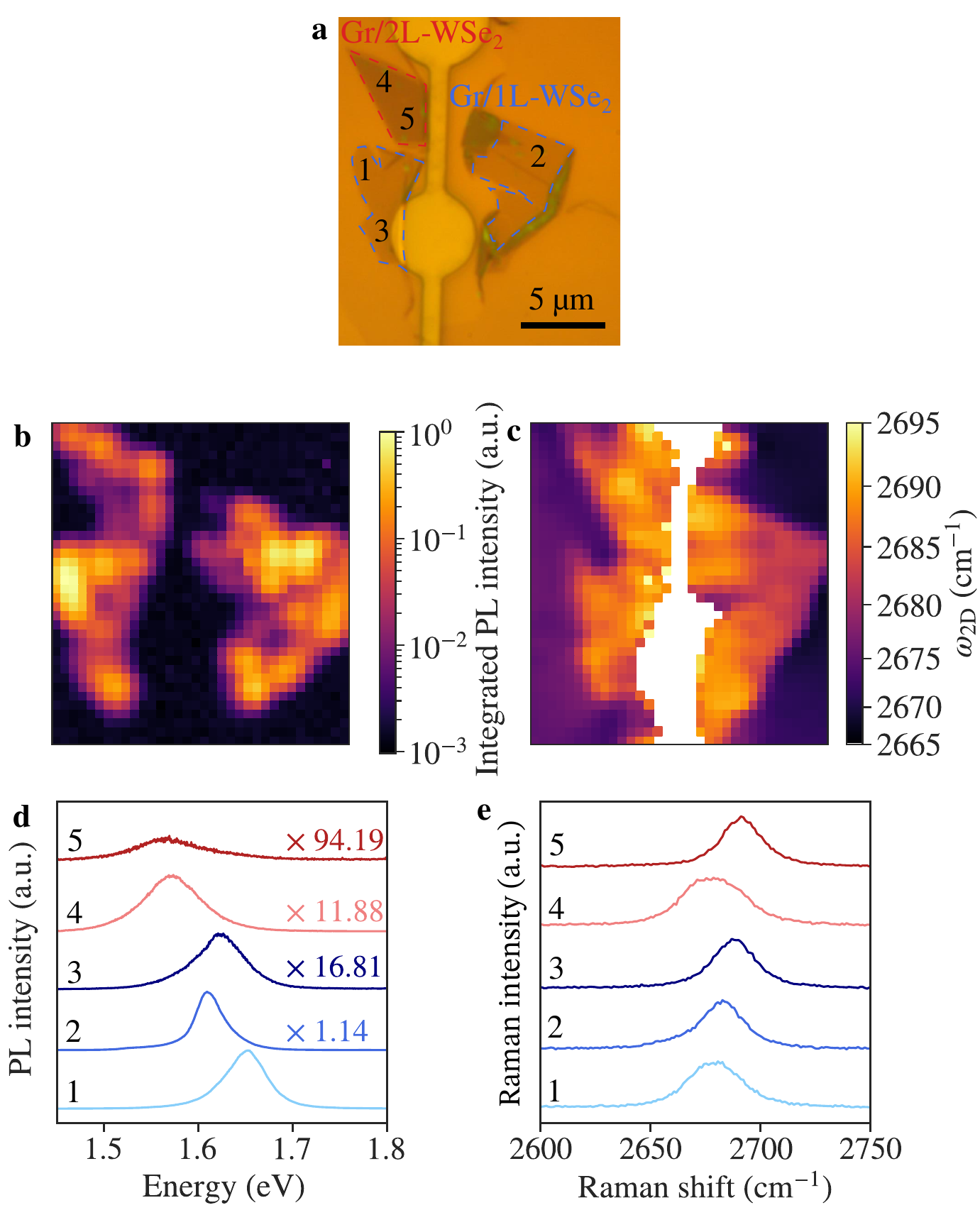}
    \caption{\textbf{Supplementary data on a spatially inhomogneous sample.} (a) Optical image of another graphene/WSe$_2$ sample on Si/SiO$_2$ with different regions of interest identified (Sample 4). (b, c) Photoluminescence intensity and 2D-mode frequency maps of the sample shown in (a, respectively. (D) Typical PL spectra for each region of interest. (E) Typical Raman 2D-mode spectra for each region of interest. Note that although the substrate is patterned with holes and venting channels, the sample is not partly suspended.}
    \label{fig_2LTMD}
    \end{center}
\end{figure}

\begin{figure}[h!]
    \begin{center}
    \includegraphics[width=0.8\linewidth]{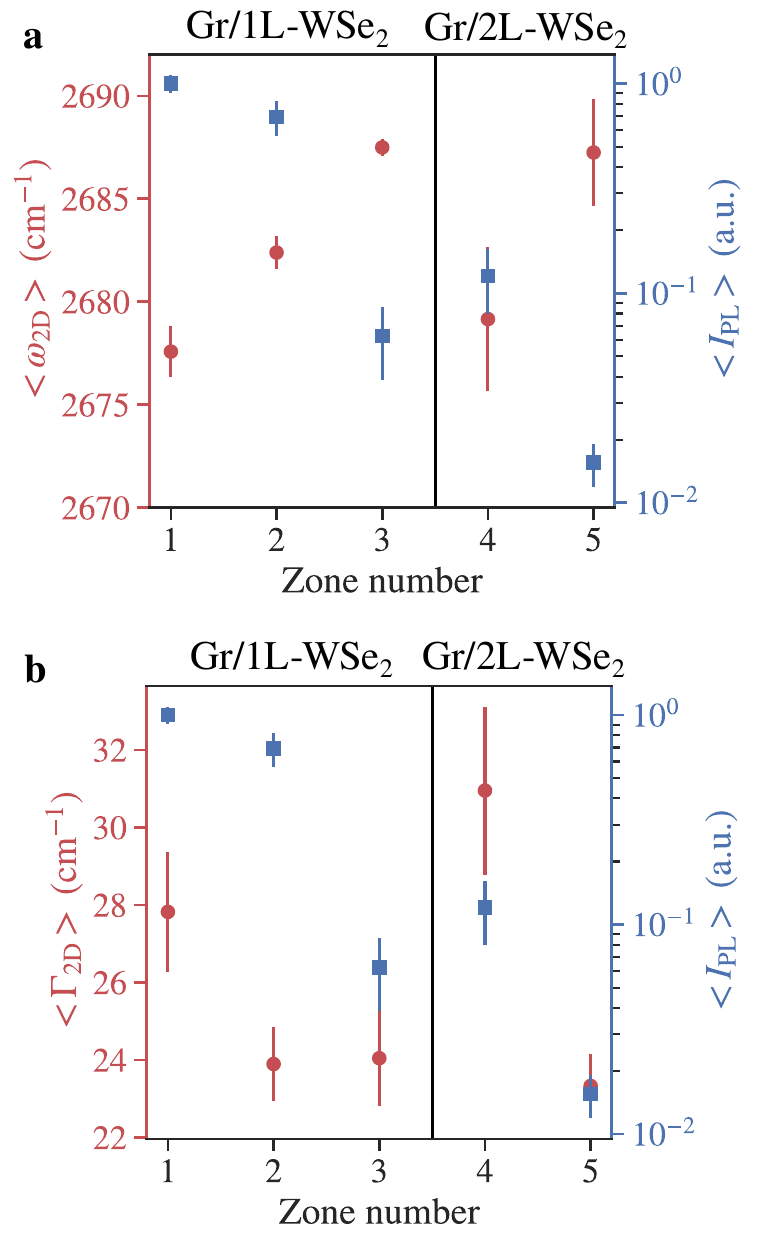}
    \caption{\textbf{Spatially averaged parameters for Sample 4.} (a) Spatially averaged 2D-mode frequency $\langle \wDD \rangle$ and integrated PL intensity $\langle I_\mathrm{PL} \rangle$ and (b) spatially averaged 2D-mode FWHM $\langle \GammaDD \rangle$ and integrated PL intensity $\langle I_\mathrm{PL} \rangle$ for the 5 different regions of interest in Sample 4, introduced in Fig.~\ref{fig_2LTMD}.}
    \label{fig_2LTMD_2}
    \end{center}
\end{figure}

Fig. \ref{fig_2LTMD}a shows an optical image of the considered sample, where we highlight five regions of interest. These regions were identified by considering the WSe$_2$ PL intensity map presented in Fig.~\ref{fig_2LTMD}b. Regions 1, 2 and 3 correspond to graphene/WSe$_2$ heterostructures with different interlayer coupling, as illustrated by their respective PL emission spectra shown Fig.~\ref{fig_2LTMD}d (see also Fig.~\ref{fig_2LTMD_2}b). The energy shift visible between region 1 and regions 2 and 3 can also be seen as an effect of dielectric screening due to a smaller interlayer distance between the two layers in regions 2 and 3 with respect to region 1. Fig.~\ref{fig_2LTMD}e and Fig.~\ref{fig_2LTMD_2}a prove the strong sensitivity of $\wDD$ to the coupling between the TMD layer and graphene, as $\wDD$ increases when the PL intensity diminishes. Note that in this specific sample, the graphene reference gives $\langle \wDD \rangle \approx 2670.5 \pm 1$ $\wn$. The sharpening of the 2D mode discussed in Appendix~\ref{section_FWHM} is also observed here depending on the interlayer coupling (see Fig.~\ref{fig_2LTMD_2}b).
Regions 4 and 5 of the same sample are formed by bilayer WSe$_2$ (2L-WSe$_2$) interfaced with graphene and present two different levels of coupling (see Fig.~\ref{fig_2LTMD}). Note that the PL intensity of bilayer WSe$_2$ is lower than in the monolayer limit because of the indirect to direct bandgap transition occurring in the monolayer limit~\cite{Zhao2013}. These two areas also show a coupling-dependent blueshift of $\wDD$ and reduction of $\GammaDD$ (see Fig.~\ref{fig_2LTMD_2}c,e). More interestingly, the obtained values of $\wDD$ and $\GammaDD$ for graphene/2L-WSe$_2$ are similar to the ones observed for graphene/1L-WSe$_2$, which suggests that dielectric screening essentially stems from the first layer in contact with graphene, highlighting the short-range character of the phenomenon.

\subsection{Impact of surface roughness}
\label{Sec_tuning_roughness}

\begin{figure}[t!]
    \begin{center}
    \includegraphics[width=1\linewidth]{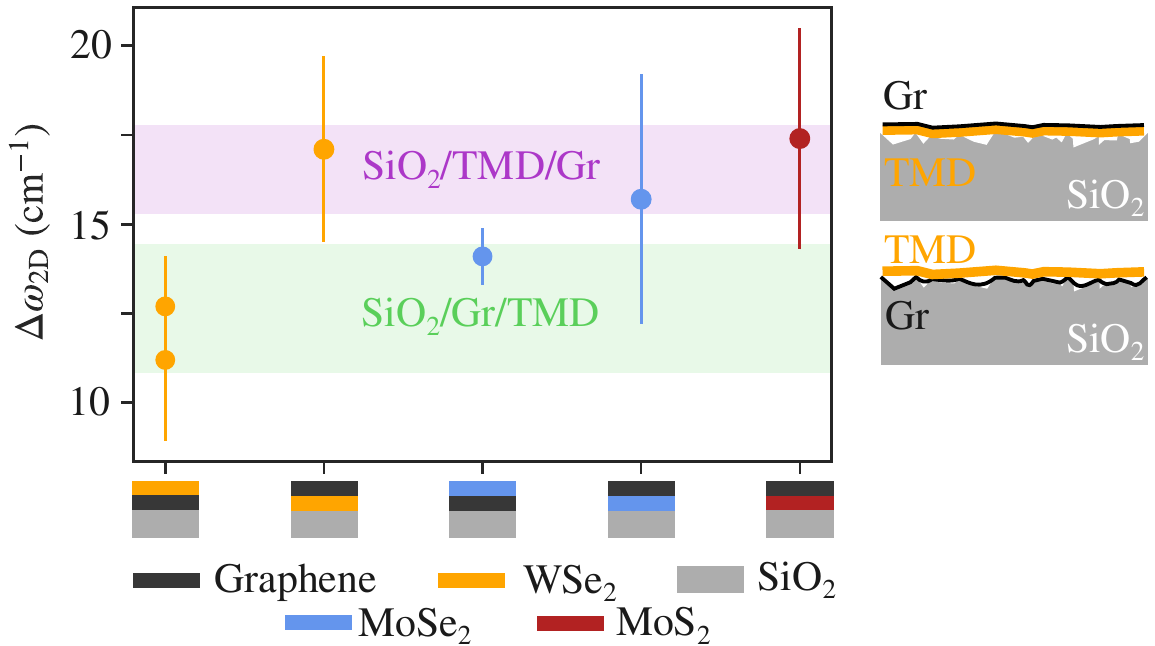}
    \caption{\textbf{Impact of surface roughness.} Spatially averaged screening-induced upshift of the 2D-mode feature in a set of graphene/TMD heterostructures deposited onto SiO$_2$, with the graphene layer either sandwiched between SiO$_2$ and the TMD or deposited on top of the SiO$_2$/TMD structure.}
    \label{fig_TMDCompar}
    \end{center}
\end{figure}
To investigate on the effect of surface roughness of the used substrate, we compare the observed upshift of the 2D mode frequency in an additional set of graphene/TMD heterostructures deposited onto SiO$_2$ substrates. 
Fig.~\ref{fig_TMDCompar} presents $\Delta \wDD$, the difference between the 2D-mode frequency in a graphene/TMD heterostructure and in the graphene reference of the same sample. We observe that $\Delta \wDD$ is appreciably higher when the graphene layer is \textit{not} directly in contact with SiO$_2$. We understand this as originating from the thicker nature of TMDs, which may conform less than graphene to the microscopic irregularities of the SiO$_2$ surface. Thus, having the TMD layer in contact with SiO$_2$ gives a flatter surface from the view point of graphene, and balancing the dielectric environment towards a stronger contribution from the TMD than in the opposite situation.

~

\subsection{Screening of phonons versus screening of excitons}
\label{Sec_tuning_PhononvsExciton}

As discussed above and in the main manuscript, the screening of optical phonons stands out as an extremely short-range effect (typically, the sub-nanometre van der Waals gap between coupled 2D layers) in contrast to the case of excitons, where longer range effects have been observed~\cite{Froehlicher2018,Tebbe2023}. We attribute this difference to the fact that for the excitons, the excited electron interacts both with the electrostatic potential of the hole as well as with the potential of the screening cloud originating from all over space, and in general from several neighboring layers. Mathematically, the effective potential seen by the electron is given in terms of the dielectric response function $\varepsilon^{-1}(\mathbf{r},\mathbf{r}')$ convoluted with the electrostatic potential of the hole, \textit{i.e.}, the dielectric response function is sampled over a large volume. By contrast, for the phonon, the interatomic force constants that define the dynamical matrix can be written as the projection of $\varepsilon^{-1}(\mathbf{r},\mathbf{r}')$ at both points $\mathbf{r}$ and $\mathbf{r}'$ onto the change of the lattice potential with respect to the ionic displacements~(cf. chap. 4 of Ref.~\cite{reichardt2018}). As the change of the lattice potential with respect to the atomic displacements falls off much faster in space (as $1/r^2$) than the electrostatic potential of the hole ($1/r$), it is much more localised and hence the dielectric response function is only sampled mostly over the layer in which the phonon resides in and its immediate neighboring layer.


\onecolumngrid

\section{Symmetry-protected zone-centre optical phonons}
\label{SectionTheory}

In this Appendix, we provide all the necessary details and definitions to understand Eqs.~(1) and (2) from the main text.
Firstly, Eq.~(1) holds for the following definition of the bare and screened electron-phonon coupling matrix elements.
The definition of the bare electron-phonon coupling matrix elements is
\begin{equation}
    g^{\q,\l;\mathrm{(bare)}}_{\k,\pi^*,\pi} \equiv \intr \, \psi^*_{\k+\q,\pi^*}(\r) \psi_{\k,\pi}(\r) \frac{\pd V_{\mathrm{lat}}}{\pd R_{\q,\l}}(\r),
\end{equation}
where $\pd/\pd R_{\q,\l} \equiv \sum_{n,\kappa,i} u^{\q,\l}_{n,\kappa,i} \pd/\pd R_{n,\kappa,i}$ is the directional derivative with respect to ionic displacements according to a phonon displacement pattern $u^{\q,\l}_{n,\kappa,i}$, where $\q$ is the phonon wave vector, $\lambda$ the phonon branch, $n$ denotes the $n$th unit cell, $\kappa$ runs over the atoms within one unit cell, and $i$ runs over the Cartesian directions. $\psi_{\k,\pi^{(*)}}(\r)$ denotes the wave function of an electron with wave vector $\k$ in the $\pi^{(*)}$-band of graphene and $V_{\mathrm{lat}}(\r)$ denotes the lattice potential.

The exact screened electron-phonon coupling matrix elements is defined as (see chap. 4 of  \cite{reichardt2018} for a derivation and motivation):
\begin{equation}
    g^{\q,\l}_{\k,\pi^*,\pi} \equiv \intr \intr' \, \psi^*_{\k+\q,\pi^*}(\r) \psi_{\k,\pi}(\r') \pd\mathcal{V}^{\q,\l}(\r,\r'),
\end{equation}
where $\pd\mathcal{V}^{\q,\l}(\r,\r')$ denotes the irreducible vertex function.
In density functional (perturbation) theory, it is just the local change of the self-consistent potential.
In general, however, it can be defined as

\begin{equation}
\begin{split}
    \pd\mathcal{V}^{\q,\l}(\r,\r')  \equiv  &  -\intt \, \frac{\pd}{\pd R_{\q,\l}}G^{-1}(\r,t;\r',0) \\
	 =&  +\intt \intr_1 \intt_1 \intr_2 \intt_2 \intr_3 \intt_3 \, G^{-1}(\r,t;\r_1,t_1) \\
       &  \quad \times  \frac{\pd V_{\mathrm{lat}}}{\pd R_{\q,\l}}(\r_3)  \Lambda(\r_3,t_3;\r_1,t_1;\r_2,t_2) G^{-1}(\r_2,t_2;\r',0),
\end{split}
\end{equation}
where $G(\r_1,t_1;\r_2,t_2) \equiv (-i) \langle 0| \mathcal{T}\{\ppsi(\r_1,t_1)\ppsidag(\r_2,t_2)\} |0 \rangle$ is the exact one-electron Green's function (we set $\hbar\equiv1$), and $\Lambda(\r_3,t_3;\r_1,t_1;\r_2,t_2)$ is the reducible vertex function defined as
\begin{equation}	
 \Lambda(\r_3,t_3;\r_1,t_1;\r_2,t_2)  \equiv \langle 0 | \mathcal{T}\left\{ \hat{\rho}(\r_3,t_3) \ppsi(\r_1,t_1) \ppsidag(\r_2,t_2) \right\} | 0 \rangle\Big|_{\mathrm{connected}}.
\end{equation}

Here, the subscript ``connected'' refers to the fully connected part of the correlation function, $\mathcal{T}$ denotes the time-ordering symbol, $\ppsi(\r,t)$ the electron field operator in the Heisenberg picture, and $\hat{\rho}(\r,t)$ the operator of the electronic charge density in the Heisenberg picture.
As alluded to earlier, we note that our definition of the screened electron-phonon coupling reduces, for instance, to the one used in density functional perturbation theory, when the corresponding one-electron Green's function is used (see chap. 4 of  \cite{reichardt2018} for a proof).\\
\\
For the special case of graphene and for the $\pi$ bands in the vicinity of the $K,K'$-points in the first Brillouin zone, we can work with the effective low-energy expansion of the electronic Hamiltonian and the electron-phonon interaction Hamiltonian, both of which are determined by the lattice symmetry~\cite{manes2007}.
In the following, we will ignore the spin degree of freedom and focus on electrons near the $K$-point.
Analogous results hold for electrons near the $K'$-point.
The low-energy effective Hamiltonian for the $\pi$-band electrons near the $K$-point reads
\begin{equation}
    \hat{H}_{\mathrm{el,eff}} = \sum_{\a,\b\in\{A,B\}} \intr \, \ppsidag_{\a}(\r) \left( -i v_{\mathrm{F}} \boldsymbol{\nabla}  \cdot \boldsymbol{\sigma}_{\a,\b} \right) \ppsi_{\b}(\r),
\end{equation}
where $v_{\mathrm{F}}$ is the exact band slope, including all screening effects from the substrate, $\boldsymbol{\sigma}$ is the vector of Pauli matrices in sublattice space, and the pseudo-spinor field operator is defined as
\begin{equation}
    \ppsi_{\a}(\r) \equiv \sum_n \frac{1}{\sqrt{N}} \psi_{p_z}(\r-\boldsymbol{\tau}_{\a}-\mathbf{R}_n) \ppsi(\r)
\end{equation}
in terms of the wave functions $\psi_{p_z}(\r)$ of the $p_z$ orbitals centered at the position $\tau_{\a}$ of the $n$th of the $N$ unit cells with origin at $\mathbf{R}_n$.

The effective bare electron-phonon interaction Hamiltonian for the $\q\to\mathbf{0}$ optical phonons ($\lambda=$LO,TO) is likewise determined by symmetry~\cite{Ando2006,manes2007} and can be written as
\begin{equation}
    \hat{H}^{\q,\l}_{\mathrm{el-ph,eff}} = F_0 \intr \, \mathrm{e}^{i\q\cdot\r} \left(\hat{\mathbf{u}}^{\q,\l}_A -\hat{\mathbf{u}}^{\q,\l}_B \right) \wedge \hat{\mathbf{J}}(\r),
\end{equation}
where $\hat{\mathbf{J}}(\r) = \sum_{\a,\b\in\{A,B\}} \ppsidag_{\a}(\r) \boldsymbol{\sigma}_{\a,\b} \ppsi(\r)$ is the electronic current operator for a Weyl fermion, $\mathbf{a} \wedge \mathbf{b} \equiv a_x b_y - a_y b_x$ the two-dimensional wedge product, and $\hat{\mathbf{u}}^{\q,\l}_{\a}$ is the vectorial lattice displacement operator for the atoms on sublattice $\a=A,B$ projected on phonon mode $(\q,\l)$.
Since this is the bare electron-phonon interaction Hamiltonian, the constant $F_0$ is by definition independent of the substrate screening.

It follows that the non-trivial part of the bare electron-phonon coupling matrix element can be written as
\begin{equation}
    g^{\q,\l;\mathrm{(bare)}}_{\k,\pi^*,\pi} \propto \sum_{\a,\b\in\{A,B\}}
    \left(\chi^{\a}_{\k+\q,\pi^*}\right)^* \chi^{\b}_{\k,\pi}  \left(\mathbf{u}^{\q,\l}_A -\mathbf{u}^{\q,\l}_B\right) \wedge \boldsymbol{\sigma}_{\a,\b},
\end{equation}
where the two-dimensional pseudo-spinors $\chi_{\k,\pi^{(*)}}$ obey the (screening-independent) Weyl equation
\begin{equation}
    \k \cdot \sum_{\b} \boldsymbol{\sigma}_{\a,\b} \chi^{\b}_{\k,\pi^{(*)}} = \substack{(+) \\ -} |\mathbf{k}| \chi^{\a}_{\k,\pi^{(*)}}.
\end{equation}

The screened electron-phonon matrix elements for the $\pi$-bands can be written in similar form:
\begin{equation}
	g^{\q,\l}_{\k,\pi^*,\pi}  \propto  \sum_{\a,\b\in\{A,B\}} \left(\chi^{\a}_{\k+\q,\pi^*}\right)^* \chi^{\b}_{\k,\pi}  \left( \mathbf{u}^{\q,\l}_A -\mathbf{u}^{\q,\l}_B \right) \wedge \boldsymbol{\pd\mathcal{V}}^{\q}_{\a,\b}(\k),
\end{equation}
with the vector- and matrix-valued irreducible vertex function 
\begin{equation}
    \boldsymbol{\pd\mathcal{V}}^{\q}_{\a,\b}(\k) \equiv \intt \intt_1 \intt_2 \sum_{\a',\b'\in\{A,B\}} G^{-1}_{\a,\a'}(\k+\q;t,t_1)
    \boldsymbol{\Lambda}^{\q}_{\a',\b'}(\k;t_1,t_2) G^{-1}_{\b',\b}(\k;t_2,0),
\end{equation}
which is in turn given in terms of the reducible vertex function
\begin{equation}
\begin{split}
    \boldsymbol{\Lambda}^{\q}_{\a',\b'}(\k;t_1,t_2) \equiv& \intr_1 \intr_2 \intt_3 \, 
    \mathrm{e}^{-i(\k+\q)\cdot\r_1}\mathrm{e}^{i\k\cdot\r_2}\\
    & \quad \times \langle 0 | \mathcal{T}\left\{ \hat{\mathbf{J}}(\boldsymbol{0},t_3) \ppsi_{\a}(\r_1,t_1) \ppsidag_{\b}(\r_2,t_2) \right\} | 0 \rangle\Big|_{\mathrm{connected}}.
\end{split}
\end{equation}
and the matrix-valued electron Green's function
\begin{equation}
    G^{-1}_{\a,\b}(\k;t_1,t_2) \equiv \int \frac{\mathrm{d}\omega}{2\pi} \mathrm{e}^{-i\omega(t_1-t_2)}
    \left[ \omega \delta_{\a,\b} - v_{\mathrm{F}} \k \cdot \boldsymbol{\sigma}_{\a,\b} +i \,\mathrm{sgn}(\omega)0^+ \right],
\end{equation}
wherein $\mathrm{sgn}$ denotes the sign function and $0^+$ a positive infinitesimal.

For the reducible vertex function, a Ward identity exists, which relates the reducible vertex function to the one-particle Green's function. Note that this Ward identity is a result of global $U(1)$ gauge invariance, or, equilvalently, the result of electron number conservation. In the limit $\q\to\mathbf{0}$, the mentioned Ward identity reads~\cite{peskin1995,srednicki2007}
\begin{equation}
    \boldsymbol{\Lambda}^{\q\to\mathbf{0}}_{\a,\b}(\k;t_1,t_2) \propto \frac{\pd}{\pd \k} G_{\a,\b}(\k;t_1,t_2).
\end{equation}
It follows immediately, that the irreducible vertex function is proportional to the momentum derivative of the inverse Green's function:
\begin{equation}
    \boldsymbol{\pd\mathcal{V}}^{\q\to\mathbf{0}}_{\a,\b}(\k) \propto \frac{\pd}{\pd \k} G^{-1}_{\a,\b}(\k;t_1,t_2) \propto v_{\mathrm{F}},
\end{equation}
which directly implies Eq.~(2) of the main text.
As mentioned in the main text, this proportionality means that all effects of screening, encoded in the exact value of the Fermi velocity, cancel out in squared phonon frequency, leading to the robustness of the G~mode frequency with respect to changes in the screening environment.

As a side point, we want to emphasize that the scaling of the screened electron-phonon coupling with the exact Fermi velocity if not obvious for optical phonons. For acoustic phonons, it is often argued~\cite{Basko2008b,Sohier2014} that a local $U(1)$ gauge symmetry exists, i.e., the action is invariant under a position-dependent change of phase of the electron field and the adding of the gradient of a position-dependent function to the phonon displacement field (akin to the standard local U(1) gauge transformation in quantum electrodynamics).
While this local $U(1)$ symmetry does indeed exist for acoustic phonons, it is not required by any more fundamental physics principle or necessary to derive the scaling of the screened electron-phonon phonon coupling. Moreover, for optical phonons, their finite frequency breaks the local $U(1)$ gauge symmetry, just as a finite photon mass would break the local gauge symmetry in electrodynamics. As such, the arguments used for the scaling of the electron-phonon coupling of acoustic phonons cannot be used in our case, requiring our more detailed theoretical reasoning.

\twocolumngrid


%

\onecolumngrid
\newpage
\begin{center}
{\Large\textbf{Supplemental Material for: Symmetry-dependent dielectric screening of optical phonons in monolayer graphene}}
\end{center}

\setcounter{equation}{0}
\setcounter{figure}{0}
\setcounter{section}{0}
\renewcommand{\thetable}{S\arabic{table}}
\renewcommand{\theequation}{S\arabic{equation}}
\renewcommand{\thefigure}{S\arabic{figure}}
\renewcommand{\thesection}{S\arabic{section}}
\renewcommand{\thesubsection}{S\arabic{section}\alph{subsection}}
\renewcommand{\thesubsubsection}{S\arabic{section}\alph{subsection}\arabic{subsubsection}}

\linespread{1.4}\selectfont

\bigskip



\section{Data analysis}
\label{SectionData}
\subsection{Fitting procedure}

\begin{figure}[h!]
    \begin{center}
    \includegraphics[width=0.7\linewidth]{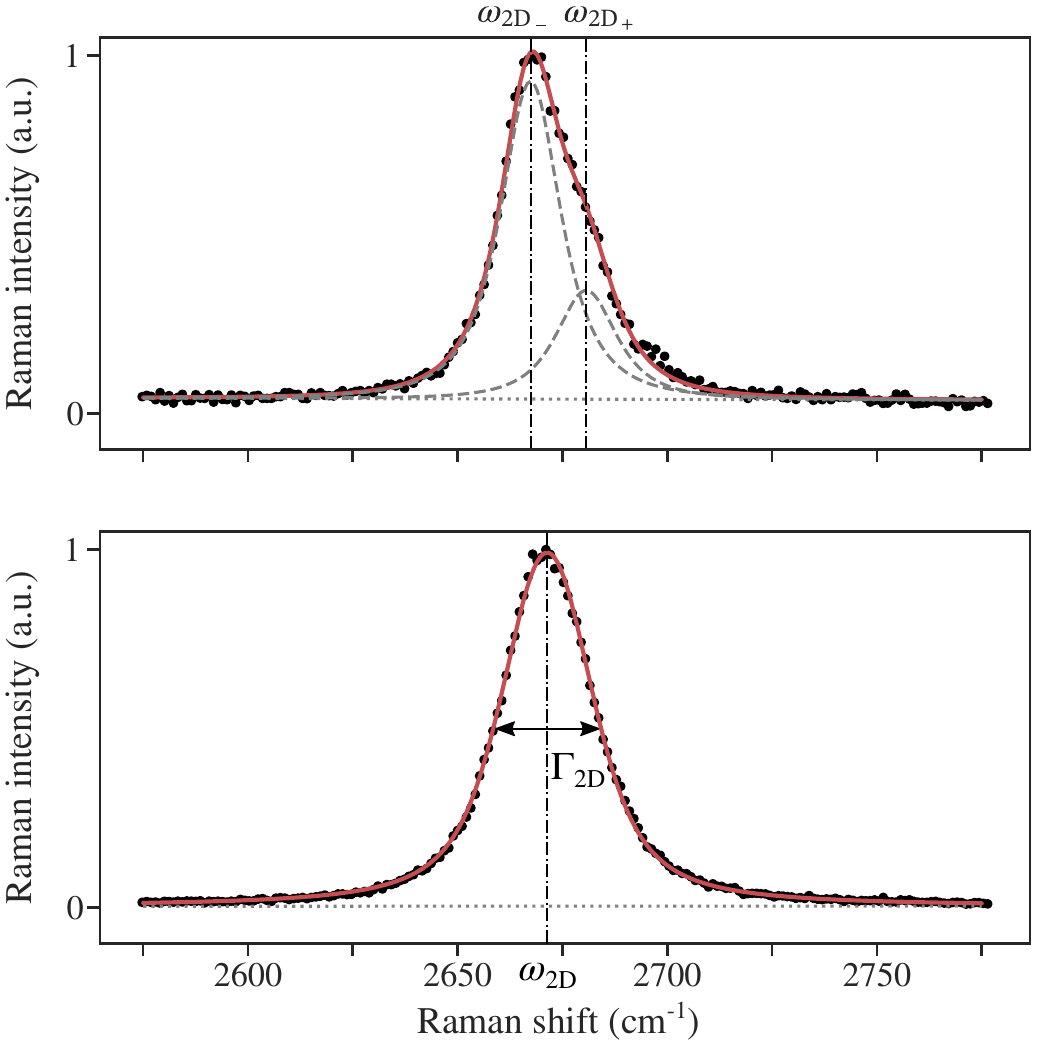}
    \caption{\textbf{Examples of fits of 2D-mode spectra.} Top: fit of the 2D-mode spectrum of suspended graphene with a double modified lorentzian profile. The two features are identified in dashed grey lines together with their corresponding central frequencies $\omega_{\mathrm{2D}_\pm}$. Bottom: fit of the 2D-mode spectrum of SiO$_2$-supported graphene with a Voigt profile.}
    \label{fig_fit}
    \end{center}
\end{figure}

To extract the relevant spectral parameters of each considered heterostructure, we followed a specific fitting procedure. For all samples, the Raman G-mode feature of graphene was fit using a Lorentzian profile to directly extract the peak position $\wG$ and full width at half maximum $\GammaG$. The 2D$^{\prime}$-mode feature was also fit to a Lorentzian profile. Except for the case of suspended graphene (see below) and unless otherwise specified, the 2D-mode feature was fit to a Voigt profile. $\GammaDD$ is then taken to be equal to the FWHM of the Voigt profile, $\wDD$ is taken as its central position. 

To determine $\wDD$ in suspended graphene, we rely on the phenomenological discussion proposed in Ref.~\onlinecite{Berciaud2013} and fit the 2D-mode spectra with a double modified Lorentzian profile~\cite{Basko2008} $f(\omega) = f_+(\omega) + f_-(\omega)$ with 
$f_{\pm}(\omega) \propto \left[ (\omega - \omega_{\mathrm{2D}_\pm})^2 
+ \frac{\gamma}{4 ( 2^{2/3} - 1 )} \right]^{-3/2}$, $\gamma$ being a common parameter for the two sub-features. Since the $\mathrm{2D}_-$ feature has much larger spectral weight than the $\mathrm{2D}_+$ feature, we define the 2D-mode frequency as $\wDD = \omega_{\mathrm{2D}_-}$. Fig.~\ref{fig_fit} presents examples of the two types of fits considered.


\subsection{Spatial averaging}

\begin{figure}[b]
    \begin{center}
    \includegraphics[width=1\linewidth]{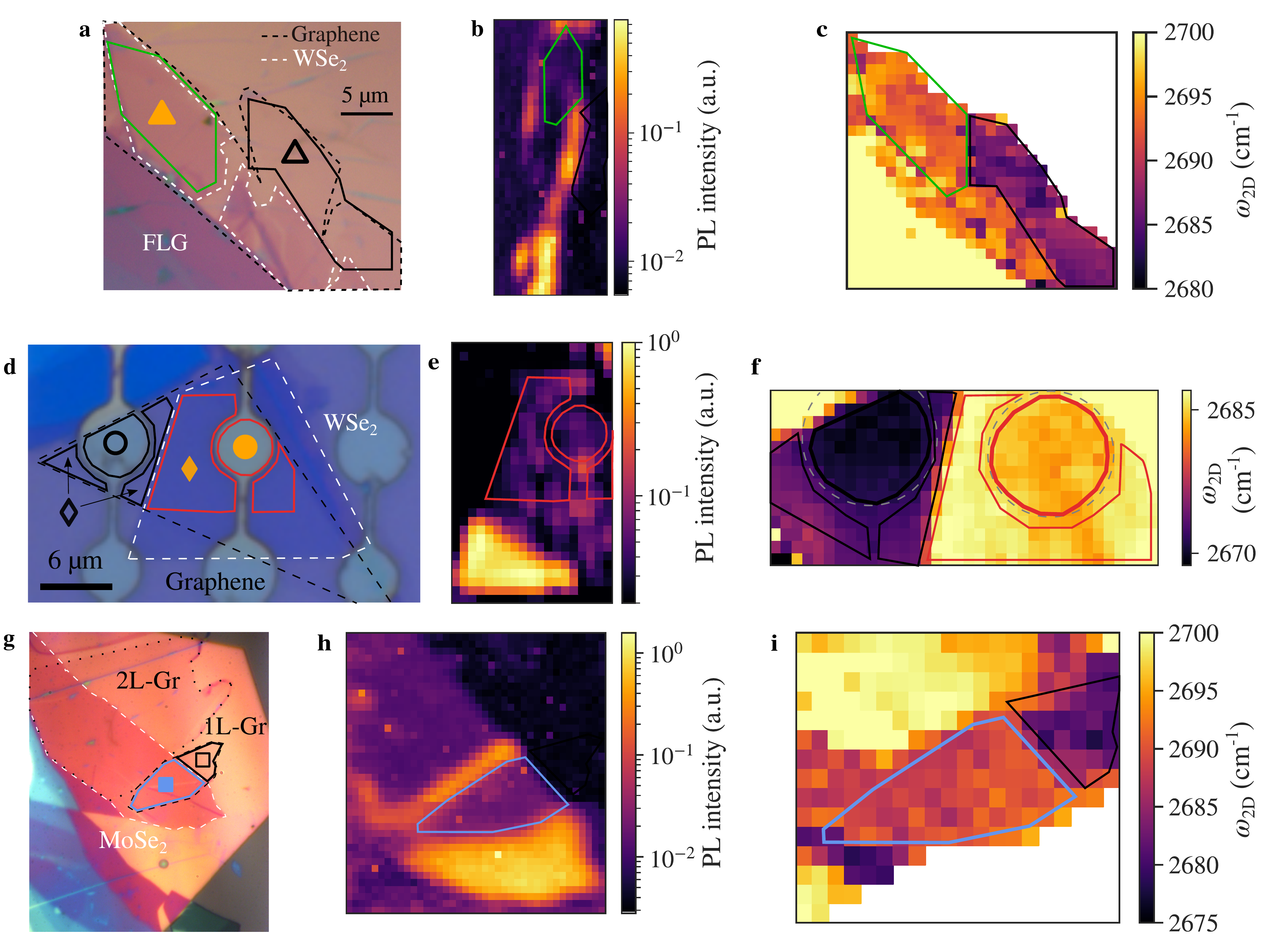}
    \caption{\textbf{Identification of the areas of interest.} Top row: Optical image, PL intensity map and $\wDD$ map of Sample 1 with hBN-capped graphene and graphene/WSe$_2$ areas. Middle row: Optical image, PL intensity map and $\wDD$ map of Sample 2 with SiO$_2$-supported graphene and graphene/WSe$_2$ areas together with their suspended counterparts. Bottom row: Optical image, PL intensity map and $\wDD$ map of Sample 3 with hBN/graphene and hBN/MoSe$_2$/graphene areas.}
    \label{fig_avg}
    \end{center}
\end{figure}

Spatial averages of the fitting parameters presented in Fig.~3 of the main text are obtained by considering a given region of interest (see Fig.~\ref{fig_avg} for a depiction of all the considered areas). The regions of interest are first identified using the optical images of our samples. Graphene/TMD areas are further identified by examining the PL intensity of the TMD layer, which is strongly quenched due to near-field transfer of TMD excitons to graphene. To ensure that only well-coupled graphene/TMD heterostructures are considered, only regions displaying a PL quenching factor above 10 at room temperature are considered.

The error bars associated with the spatially averaged values are given by the standard deviation of each parameter in a specific area. 

\newpage
\section{Determining built-in strain and residual doping}
\label{SectionStrainDoping}
\subsection{Built-in strain and unintentional doping}

\begin{figure}[h!]
    \begin{center}
    \includegraphics[width=1\linewidth]{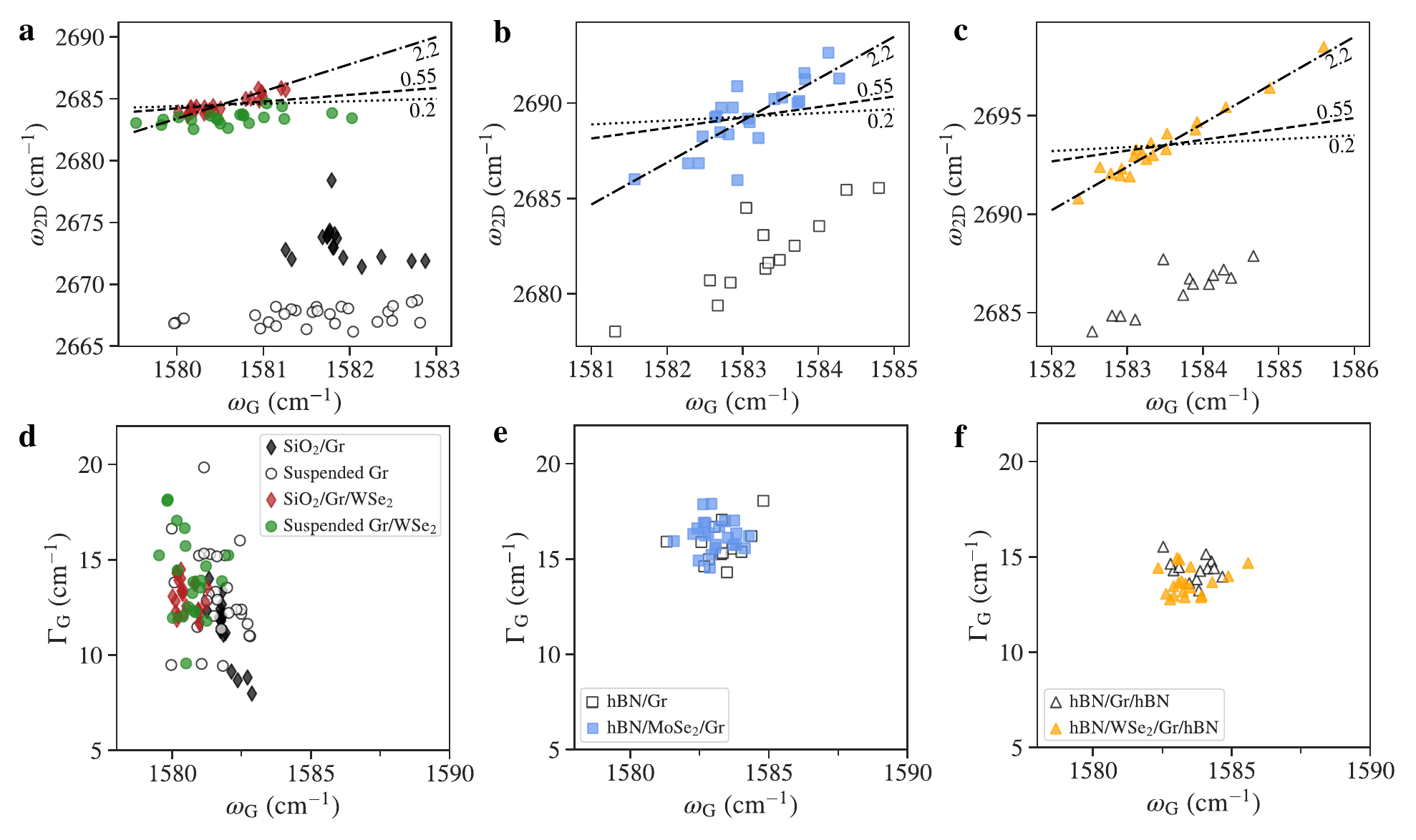}
    \caption{\textbf{Correlation plots.} (a, b, c) $\wDD$ vs $\wG$ correlations for the three samples shown in Fig. 3 of the main text (Samples 2, 3 and 1 in Fig.~\ref{fig_avg}, respectively). The plotted lines give the expected correlation slopes for electron doping ($\partial \wDD / \partial \wG \approx 0.2$), hole doping ($\partial \wDD / \partial \wG \approx 0.55$) and strain ($\partial \wDD / \partial \wG \approx 2.2$). (d, e, f) $\GammaG$ versus $\wG$ correlations for the same samples.}
    \label{fig_strain}
    \end{center}
\end{figure}

For each type of heterostructure, we examine the $\wDD$ versus $\wG$ and $\GammaG$ versus $\wG$ correlations, as shown in Fig. \ref{fig_strain}. As extensively documented elsewhere~\cite{Lee2012,Metten2013,Metten2014,Neumann2015,Froehlicher2015}, such plots provide quantitative information on the residual charge carrier density (or non-intentional doping) and built-in strain field of the considered graphene area. First, the $\GammaG$-$\wG$ correlation allows to estimate the doping level based on robust models for doping induced zone-centre optical phonon renormalisation~\cite{Ando2006,Yan2007,Pisana2007}. For all cases, we find spatially averaged values $\langle \GammaG \rangle > 10$ $\wn$ and no clear $\wG$-$\GammaG$ correlation (see Fig. \ref{fig_strain}D-F and Fig.~5 in the main text). These observations indicate that unintentional doping leads to Fermi level shifts typically less than 100~meV away from the Dirac point~\cite{Froehlicher2015}. 

The $\wDD$-$\wG$ correlation allows us to separate the effect of doping and strain and hence resolve small variations of built-in strain in the studied hBN-based heterostructures (see Fig.~\ref{fig_strain}b,c and main text for details). These variations lead to a scatter in the measured values of $\wDD$ and hence to slightly increased errorbars in the precise determination of the blueshift of $\wDD$ due to dielectric screening. However, despite these small variations, the $\wDD$-$\wG$ correlations differs only by a rigid (screening-induced) shift of $\wDD$ when comparing bare graphene to graphene/TMD heterostructures, consolidating the hypothesis of low built-in strain (typically $\sim 10^{-2}\,\%$) in all the samples investigated here.  

Let us finally stress that in the presence of an inhomogeneous dielectric environment, our demonstration that the correlation between the 2D- and G-mode frequencies is essentially a vertical line (see Fig. 3 in the main text) enables a straightforward identification of domains associated with distinct dielectric environments within a given sample. As illustrated in Fig.~\ref{fig_strain}a-c, these domains appear as vertically offset clouds of points in the $(\wG,\wDD)$ plane. Note that such a vertical offset in the $(\wG,\wDD)$ plane would only be only achievable by a combination of unrealistically large strain and doping levels and thus is a unique signature of a change of dielectric environment. At the same time, the magnitude of these vertical shifts is virtually independent of the level of strain and doping. In each of these domains, some scattering may be seen due to nanoscale strain and doping gradients that follow the well-known correlations in the $(\wG,\wDD)$ plane~\cite{Lee2012a,Froehlicher2015,Metten2014,Neumann2015}, allowing a quantitative determination of the residual charge carrier density and strain level within an identified dielectric environment. Such a multiparameter analysis can be further consolidated using the analysis of $\GammaG$ (Fig.~\ref{fig_strain}d-f) and of the integrated intensity ratio between the 2D-mode and G-mode intensities (Sec.~\ref{Sec_Int} and Fig.~\ref{fig_Raman_Int}).

\subsection{Photoinduced doping}

\begin{figure}[h!]
    \begin{center}
    \includegraphics[width=1\linewidth]{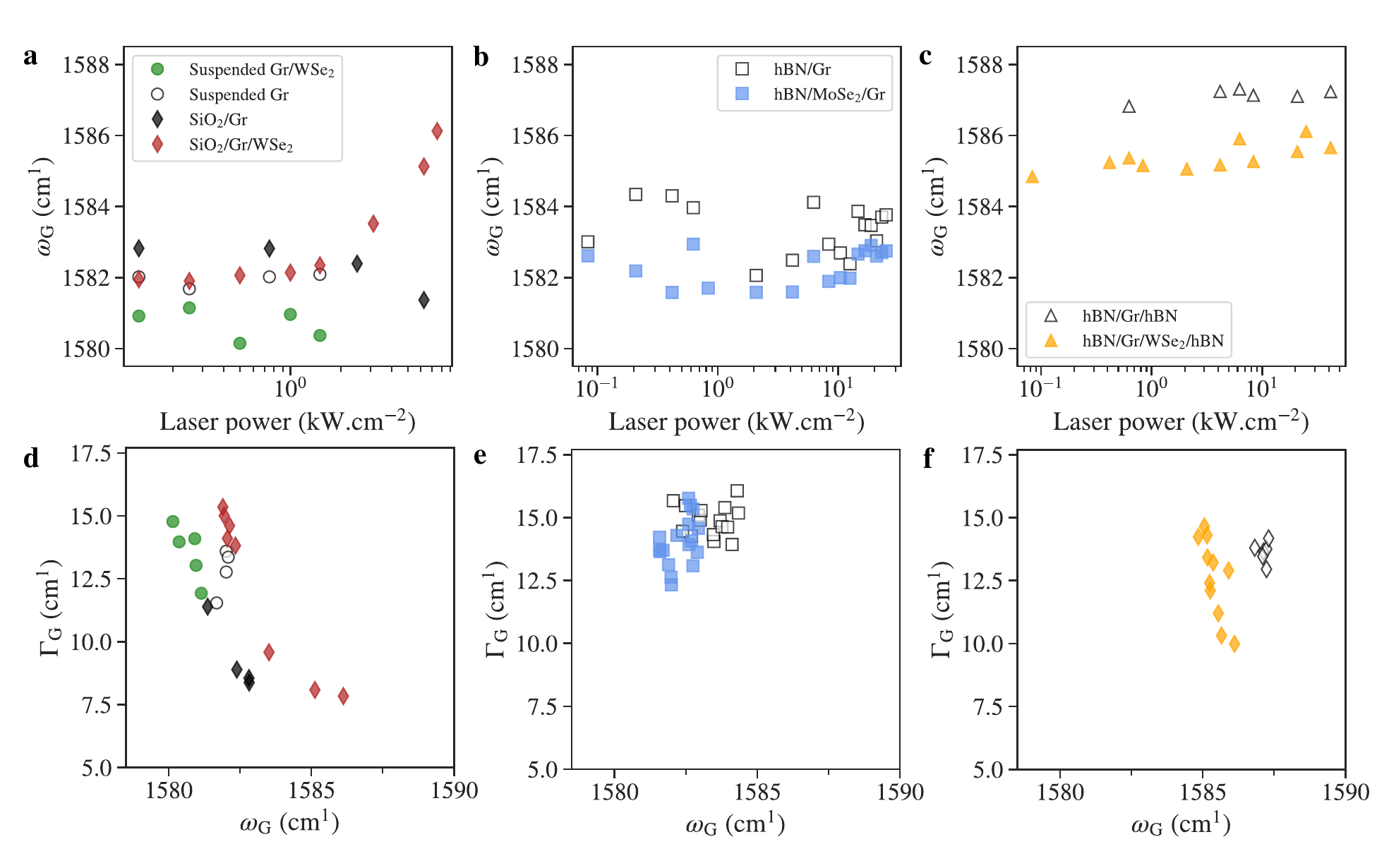}
    \caption{\textbf{Assessing photoinduced doping.} Evolution of $\wG$  with increasing laser intensity  (a, b, c) and $\GammaG$ versus $\wG$ correlation under increasing laser intensity (D, E, F) for Sample 2, 3 and 1, respectively (see Fig.~\ref{fig_avg}). The laser photon energy is 2.33~eV.}
    \label{fig_PRX}
    \end{center}
\end{figure}

Interaction with polar substrates and molecular adsorbates~\cite{Ryu2010} as well as interlayer charge transfer between TMD and graphene can lead to photoinduced doping of the graphene layer~\cite{Froehlicher2018} and experimentally translate into a sizeable upshift and narrowing of the G-mode feature, which may affect our results. Because of this, we performed laser intensity-dependent measurements on our samples to determine the threshold of such unwanted photoinduced doping. As shown in Fig.~\ref{fig_PRX}, photodoping is observable in SiO$_2$-supported samples and is negligible in suspended graphene and in hBN-supported or hBN-capped samples~\cite{parralopez2021}. All the data presented in the main text were acquired with laser intensities low enough to avoid any spurious photoinduced doping of the graphene layer (typically $\sim 100~ \mu \rm W/ \mu \rm m^{2}$).

\newpage

\section{Analysis of the Integrated Raman intensities} \label{Sec_Int}

In this section, we experimentally address the impact of dielectric screening on the integrated intensities of the Raman G- and 2D-modes, denoted $I_{\rm G}$ and $I_{\rm{2D}}$, respectively. As discussed in the main manuscript and in Appendix E, the electron-phonon coupling strength (and hence the Raman intensities) are directly affected by screening. As a logical consequence, the strong effect of dielectric screening on the TO phonon dispersion at and near $\mathbf{K}$ could, at least on the qualitative level, result in a reduction of the 2D-mode intensity. In order to test this scenario experimentally, we have carefully analysed $I_{\rm G}$ and $I_{\rm{2D}}$ in the sample, discussed in Fig.~2 of the main manuscript (graphene with and without a WSe$_2$ capping layer, with suspended and supported areas), where the effects of screening can be seen most prominently. Let us recall that in this sample, as in all the samples investigated here, the impact of unintentional doping and built-in strain on the Raman features is negligible (Sec.~\ref{SectionStrainDoping}) and, in particular, a reduction of the $I_{\rm{2D}}/I_{\rm{G}}$ ratio in WSe$_2$/graphene due to doping~\cite{Basko2009,Froehlicher2015} can safely be ruled out.

\begin{figure}[h!]
    \begin{center}
    \includegraphics[width=0.95\linewidth]{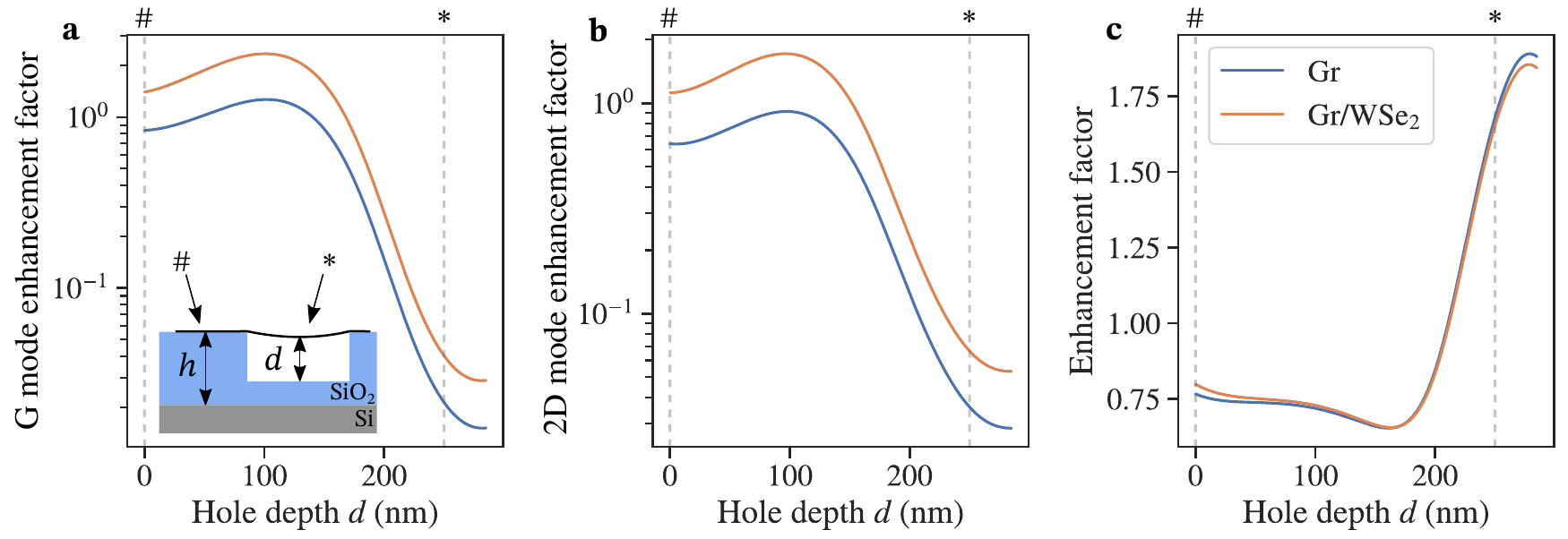}
    \caption{\textbf{Optical interference effects.} Computed interference enhancement factors for the G-mode intensity $I_{\rm{G}}$ (a), the 2D-mode intensity $I_{\rm{2D}}$ (b) and the $I_{\rm{2D}}/I_{\rm{G}}$ ratio. The geometry considered is indicated in the inset of panel (a). Specifically, the SiO$_2$ thickness of our sample is $h=285~\rm{nm}$ and the supported and suspended geometries are highlighted with $\#$ and $\ast$ symbols, respectively. The suspended geometry in our experiments corresponds to a hole depth of $d=250~\rm{nm}$. The cases of bare graphene (blue lines) and WSe$_2$/graphene (orange lines) are considered.}
    \label{fig_interference}
    \end{center}
\end{figure}

\begin{figure}[h!]
    \begin{center}
    \includegraphics[width=0.95\linewidth]{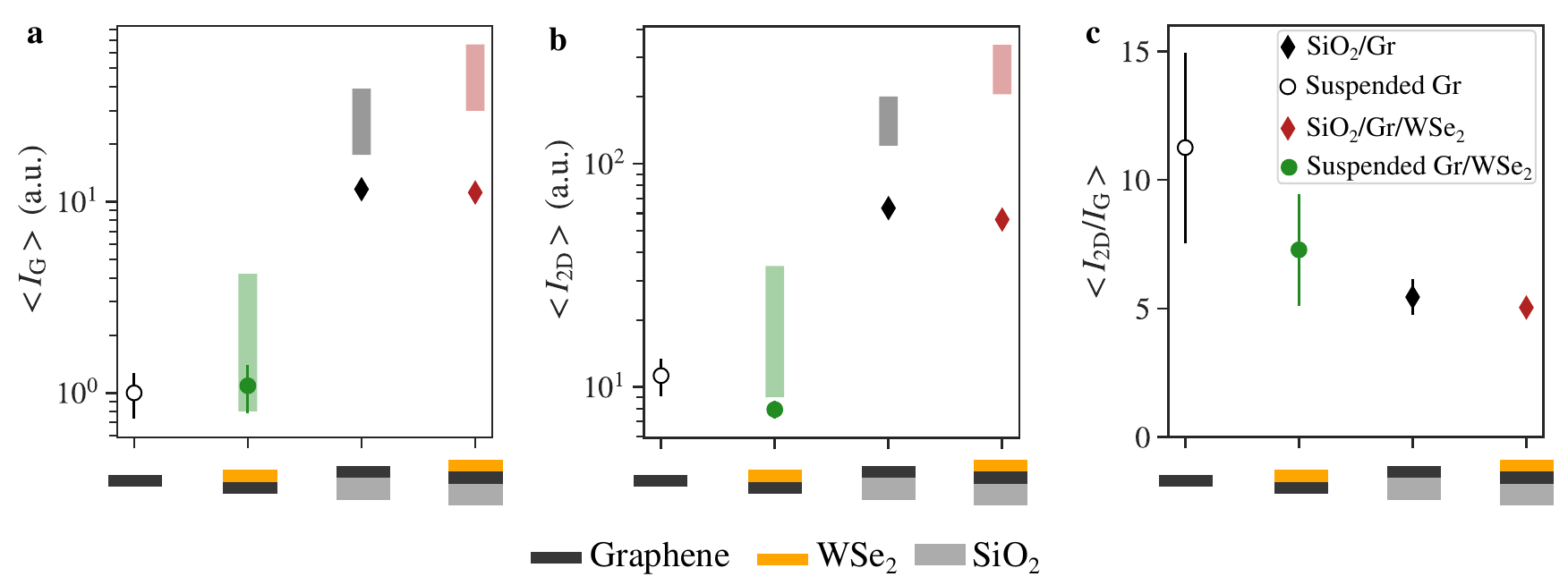}
    \caption{\textbf{Spatially averaged Raman intensities.} Spatially averaged values of the integrated Raman intensities $\langle I_{\rm G} \rangle$ (a), $\langle I_{\rm{2D}} \rangle$ (b) and integrated intensity ratio $\langle I_{\rm{2D}}/I_{\rm{G}} \rangle$ (c) for the Raman map shown in Fig.~2a of the main manuscript. The data in (a) and (b) are normalised with respect to the value of $\langle I_{\rm G} \rangle$ in bare suspended graphene. The semi-transparent vertical bars correspond to the computed values based solely on optical interference effects (see Fig.~\ref{fig_interference}). The height of the bars accounts for the experimental uncertainty in the hole depth $d$ and for a possible slack of the suspended graphene and WSe$_2$/graphene membranes (see inset in Fig.~\ref{fig_interference}a).}
    \label{fig_Raman_Int}
    \end{center}
\end{figure}

Additionally, optical interference effects due to multiple reflections in our layered structures (Fabry-Perot effects) are well-known to affect the intensity of the Raman scattering features and must be taken into account properly~\cite{Yoon2009,Metten2014,Metten2017}. Figure~\ref{fig_interference} shows the computed Raman intensity enhancement factors solely due to optical interference for TMD/graphene versus bare graphene, within the geometry indicated in the inset of Fig.~\ref{fig_interference}a. The main effects due to optical interference are the following:
\begin{itemize}
    \item The Raman intensities of the G- and 2D-modes are expected to be significantly larger, by more than one order of magnitude, in the supported regions (marked with $\#$ in Fig.~\ref{fig_interference}), compared to the suspended regions (marked with $\ast$ in Fig.~\ref{fig_interference}). 
    \item The presence of a WSe$_2$ monolayer should lead to a slight increase of the Raman intensities by a bit less than a factor of 2 (Fig.~\ref{fig_interference}a,b).
    \item $I_{\rm{2D}}/I_{\rm{G}}$ is expected to be larger in the suspended regions than in the supported regions (Fig.~\ref{fig_interference}c). This is a purely optical effect due to the fact that the enhancement factor of $I_{\rm G}$ reaches lower values than that of $I_{\rm{2D}}$ in the suspended areas, because the wavelength of the outgoing photons for the G-mode is shorter than those of the 2D-mode photons~\cite{Metten2015}.
    \item Only a marginal increase of $I_{\rm{2D}}/I_{\rm{G}}$ by a few percent is expected in TMD/graphene  compared to bare graphene (see (Fig.~\ref{fig_interference}c). This is below the standard deviation in the spatially averaged values measured over a given region of interest (Fig.~\ref{fig_Raman_Int}).
\end{itemize}

The spatially averaged values $\langle I_{\rm G} \rangle$, $\langle I_{\rm{2D}} \rangle$ and $\langle I_{\rm{2D}}/I_{\rm{G}} \rangle$ for the map shown in Fig.~2a of the main text (also shown in Fig.~\ref{fig_avg}f) are shown in Figure~\ref{fig_Raman_Int} together with the standard deviations as errorbars. For convenience, we have normalised the Raman intensities to the value of $\langle I_{\rm G} \rangle$ measured on bare suspended graphene. First and foremost, our experimental observations are in very good qualitative agreement with the trends predicted by our interference calculations. Second, we note a remarkable reduction of $\langle I_{\rm{2D}}/I_{\rm{G}} \rangle$  by a factor of approximately 1.5 when moving from suspended graphene to suspended WSe$_2$/graphene. A qualitatively similar, yet smaller reduction of 10$\%$ is also observed when comparing WSe$_2$/graphene on SiO$_2$ and graphene on SiO$_2$. We also note that these changes arise from a decrease in the 2D-mode intensity, whereas the G-mode intensity remains essentially constant when moving from graphene to WSe$_2$/graphene. These changes can tentatively by attributed to dielectric screening.

To close this section, we wish to stress that from a theoretical standpoint, it is by no means trivial to compute the changes in the absolute Raman scattering intensity of the G- and 2D-modes as a function of screening. Indeed, the Raman processes involved in these two modes are very different and may be affected by other effects, notably resonance effects and quantum interference effects~\cite{Basko2008,Basko2009b,Venezuela2011,Chen2011}, that are not directly related to dielectric screening and may overshadow the sole contribution of the dielectric screening. A comprehensive theoretical analysis of the Raman susceptibility of graphene in various dielectric environments goes well-beyond the scope of the present study.

\center\rule{8cm}{1pt}

\end{document}